\documentclass[usenatbib]{mnras}

\usepackage{graphicx}    
\usepackage{hyperref}
\hypersetup{colorlinks=true, citecolor=blue, linkcolor=blue}

\def\aj{Astron.\ J. }
\def\apj{Astrophys.\ J. } 
\def\apjl{Astrophys.\ J. }
\def\apjs{Astrophys.\ J.\ (Supp.) }
\def\aap{Astron.\ Astrophys. }

\def\gjras{Geophys.\ J.\ R.\ Astron.\ Soc. }

\def\mnras{Mon.\ Not.\ R.\ Astron.\ Soc. }

\def\nat{Nature }

\def\prd{Phys.\ Rev.\ D }

\def\rg{Rev.\ Geophys. }

\def\figpath{}
\def\bfx#1{{#1}}

\begin{document}

\title[CoRoT-7 two-planet system with a Maxwell rheology]
{Coupled orbital and spin evolution of the CoRoT-7 two-planet system using a Maxwell viscoelastic rheology}
\author[A. Rodr\'iguez et al.]{A.~Rodr\'iguez$^1$, N. Callegari Jr.$^2$, and A. C. M Correia$^{3,4}$\\
$^1$ Universidade Federal do Rio de Janeiro, Observat\'orio do Valongo, Ladeira do Pedro Ant\^onio 43, 20080-090, Rio de Janeiro, Brazil\\
$^2$ Instituto de Geoci\^encias e Ci\^encias Exatas, Unesp-Univ Estadual Paulista , Av. 24-A, 1515, 13506-900, Rio Claro, SP, Brazil\\
$^3$ CIDMA, Departamento de F\'isica, Universidade de Aveiro, Campus de Santiago, 3810-193 Aveiro, Portugal\\
$^4$ ASD, IMCCE-CNRS UMR8028, Observatoire de Paris, 77 Av. Denfert-Rochereau, 75014 Paris, France}

\date{\today}

\maketitle

\begin{abstract}
We investigate the orbital and rotational evolution of the CoRoT-7 two-planet system, assuming that the innermost planet behaves like a Maxwell body. 
We numerically resolve the coupled differential equations governing the instantaneous deformation of the inner planet together with the orbital motion of the system. 
We show that, depending on the relaxation time for the deformation of the planet, the orbital evolution has two distinct behaviours:
for relaxation times shorter than the orbital period, we reproduce the results from classic tidal theories, for which the eccentricity is always damped. 
However, for longer relaxation times, the eccentricity of the inner orbit is secularly excited and can grow to high values.
This mechanism provides an explanation for the present high eccentricity observed for CoRoT-7\,b, as well as for other close-in super-Earths in multiple planetary systems.
\end{abstract}

\begin{keywords}
Planets and satellites: dynamical evolution and stability. Planet-star interactions
\end{keywords}


\section{Introduction}\label{intro}

Close-in planets undergo tidal interactions with the central star, which shrink and circularize the orbits on time-scales that depend on the orbital distances, but also on the physical properties of the interacting bodies.
The rotation of short-period planets is also modified and reaches a stationary 
value in time-scales usually much shorter than the orbital evolution \citep[e.g.,][]{Hut_1981, Ferraz-Mello_etal_2008, Correia_2009, Rodriguez_etal_2011}. 
The tidal interaction ultimately results in synchronous motion (the orbital and rotation periods become equal), \bfx{which is the 
only possible} state when the orbit is circularized \citep[e.g.,][]{Hut_1981, Ferraz-Mello_etal_2008}.
However, as long as the orbit has some eccentricity, the rotation can \bfx{stay} in non-synchronous configurations. 
In general, planets with a \bfx{primarily rocky} composition have a permanent equatorial deformation or frozen-in figure \citep[e.g.,][]{Goldreich_Peale_1966, Greenberg_Weidenschilling_1984}, which contributes with a conservative restoration torque \bfx{on their figures}.
In the context of the two-body problem, the gravitational interaction of an asymmetric planet with the star drives the planet rotation into different regimes of motion, including
oscillations around exact spin-orbit resonances (SOR). 
When dissipative effects are taken into account, the oscillations are damped and the planet rotation can be trapped in exact resonance \citep[e.g.,][]{Goldreich_Peale_1966, Correia_Laskar_2009}. 

Although the orbital and spin evolution are connected through the total angular momentum conservation, they are commonly studied separately due to the different time-scales involved in their evolution.
However, it has been shown that for close-in planets the tidal evolution of the coupled orbit-rotation is important and should not be disassociated \citep{Correia_etal_2012, Correia_etal_2013, Rodriguez_etal_2012, Rodriguez_etal_2013, Greenberg_etal_2013}. 
All studies cited above assumed simplified tidal models, usually using constant or linear tidal deformations \citep[e.g.,][]{Darwin_1880, Mignard_1979}, for which the tidal dissipation is constant or proportional to the corresponding frequency of the perturbation. 
A more realistic approach to deal with the dependency of the phase lag with the tidal frequency is to assume a viscoelastic rheology \citep[e.g.,][]{Efroimsky_2012, Remus_etal_2012b, Ferraz-Mello_2013, Correia_etal_2014}.
These rheologies have been shown to reproduce the main features of tidal dissipation \citep[for a review of the main viscoelastic models see][]{Henning_etal_2009}.
One of the simplest models of this kind is to consider that the planet behaves like a Maxwell material\footnote{The Maxwell material is represented by a purely viscous damper and a purely elastic spring connected in series \citep[e.g.,][]{Turcotte_Schubert_2002}.}.
In this case, the planet can respond as an elastic solid or as a viscous fluid, depending on the frequency of the perturbation.

\citet{Correia_etal_2014} studied the orbital and rotational evolution of a single close-in planet using a Maxwell viscoelastic rheology.
However, instead of decomposing the tidal potential in an infinite sum of harmonics of the tidal frequency \citep[e.g.,][]{Kaula_1964, Mathis_Poncin-Lafitte_2009, Efroimsky_2012}, they compute the instantaneous deformation of the planet using a differential equation for its gravity field coefficients.
They have shown that when the relaxation time of the deformation is larger than the orbital period (which is likely the case for rocky planets), spin-orbit equilibria arise naturally at half-integers of the mean motion, without requiring to take into account the permanent equatorial deformation.

The method by \citet{Correia_etal_2014} has several advantages for studying the tidal evolution of planetary systems:
1) it works for any kind of perturbation, even for the non-periodic ones (such as chaotic motions or transient events);
2) the model is valid for any eccentricity and inclination value, we do not need to truncate the equations of motion;
3) it simultaneously reproduces the deformation and the dissipation on the planet.
Therefore, this model seems to be the most appropriate to also study the impact of gravitational perturbations of companion bodies in the orbit of the inner planet.
Indeed, we show here that the eccentricity of the inner body can increase due to a feedback mechanism between the tidal deformation of the planet and the orbital forcing.

In this paper we provide a simple  model for the coupled orbital and spin
evolution of an exoplanet with a companion (Sect.\,\ref{secmodel}),  
and apply it to the CoRoT-7 planetary system (Sect.\,\ref{corot}). 
We then give an explanation for the \bfx{non-zero presently observed} eccentricity values (Sect.\,\ref{pumping}),
and derive some conclusions (Sect.\,\ref{disc}).

\section{Model}
\label{secmodel}

We consider a system consisting of a central star with mass $m_0$, and two companion planets with masses $m_1$ and $m_2$, such that $m_1, m_2 \ll m_0$.
The subscript 1 always refers to the inner planet, while the 2 refers to the outer one.

The inner planet is considered an oblate ellipsoid with gravity field coefficients given by $J_2$, $C_{22}$ and $S_{22}$, whereas the star and the outer planet are considered as point masses.
We also assume that the spin axis of the inner planet, with rotation rate $\Omega$, is along the axis of maximal inertia $\vec{k}$ (gyroscopic approximation), and that $\vec{k}$ is orthogonal to its orbital plane (which corresponds to zero obliquity).
The ellipsoid can be deformed by self rotation and tidal interactions with the central star, and we adopt a Maxwell viscoelastic rheology to model the deformation of the planet \citep[see][]{Correia_etal_2014}.

\subsection{Equations of motion}

The equations of motion governing the orbital evolution of the system in a astrocentric frame are

\begin{eqnarray}\label{mov1}
\ddot{\vec{r}}_1&=&-\frac{\mu_1}{r_1^3}\vec{r}_1+Gm_2\left(\frac{\vec{r}_2-\vec{r}_1}{|\vec{r}_2-\vec{r}_1|^3}-\frac{\vec{r}_2}{r_2^3}\right)\nonumber\\
&&+\vec{f} +\vec{g}_1 +\frac{G m_2}{\mu_2} \vec{g}_2  \ ,
\end{eqnarray}
\begin{eqnarray}\label{mov2}
\ddot{\vec{r}}_2&=&-\frac{\mu_2}{r_2^3}\vec{r}_2+Gm_1\left(\frac{\vec{r}_1-\vec{r}_2}{|\vec{r}_1-\vec{r}_2|^3}-\frac{\vec{r}_1}{r_1^3}\right) \nonumber\\
&&+\vec{g}_2 +\frac{G m_1}{\mu_1} \left(\vec{f}+\vec{g}_1\right) \ ,
\end{eqnarray}
where $G$ is the gravitational constant, $\mu_i = G(m_0+m_i)$,  and $\vec{r}_i$ is the position of the planet with respect to the star (with $i=1,2$).
$\vec{g}_i$ are the additional accelerations due to general relativity corrections to the first order in $m_i/m_0$, given by \citep[see][]{Kidder_1995}
\begin{equation}\label{frel}
\vec{g}_i=-\frac{\mu_i}{c^2r_i^3}\left[\left(\dot{\vec{r}}_i\cdot\dot{\vec{r}}_i-4\frac{\mu_i}{r_i}\right)\vec{r}_i -4(\vec{r}_i\cdot\dot{\vec{r}}_i) \dot{\vec{r}}_i \right],
\end{equation}
where $c$ is the speed of light.
$\vec{f}$ is the acceleration arising from the potential created by the deformation of the inner planet, 
which is given by \citep{Correia_etal_2014}
\begin{eqnarray}\label{fdef}
\vec{f}&=&-\frac{3\mu_1R^2}{2r_1^5}J_2\vec{r}_1 -\frac{9\mu_1R^2}{r_1^5}\left[C_{22}\cos2\gamma-S_{22}\sin2\gamma\right]\vec{r}_1\nonumber\\
&&+\frac{6\mu_1R^2}{r_1^5}\left[C_{22}\sin2\gamma+S_{22}\cos2\gamma\right]\vec{k}\times\vec{r}_1,
\end{eqnarray}
where $R$ is the mean radius of the inner planet, and $\gamma=\theta-\ell$, with $\theta$ the rotation angle ($\Omega=\dot{\theta}$), $\ell = \varpi + v$ the true longitude, $\varpi$ the longitude of the pericenter, and $v$ the true anomaly.


The torque acting to modify the inner planet rotation is given by

\begin{equation}\label{torque}
\ddot{\theta}=-\frac{6Gm_0m_1R^2}{Cr_1^3}\left[C_{22}\sin2\gamma+S_{22}\cos2\gamma\right], 
\end{equation}
where $C$ is the principal moment of the inertia along the axis $\vec{k}$.

The inner planet is deformed under the action of self rotation and tides. 
Therefore, the gravity field coefficients can change with time
as the shape of the planet is continuously adapting to the equilibrium figure. 
According to the Maxwell viscoelastic rheology, the deformation law for these coefficients is given by \citep{Correia_etal_2014}

\begin{eqnarray}\label{max1}
&&J_2+\tau\dot{J}_2 = J_2^0 + J_2^{r} + J_2^t \ ,\nonumber\\
&&C_{22}+\tau\dot{C}_{22}  = C_{22}^0 + C_{22}^t \ ,\\
&&S_{22}+\tau\dot{S}_{22} = S_{22}^t \ ,\nonumber
\end{eqnarray}
where $\tau$ is the relaxation time of the planet in response to deformation\footnote{$\tau = \tau_v + \tau_e$, where $\tau_v$ and $\tau_e$ are the viscous (or fluid) and Maxwell (or elastic) relaxation times, respectively. For simplicity, in this paper we consider $\tau_e=0$, since this term does not contribute to the tidal dissipation \citep[for more details, see][]{Correia_etal_2014}. Our model is thus also equivalent to a Newtonian creep model \citep{Ferraz-Mello_2013}.}.
$J_2^0$ and $C_{22}^0$ are permanent values of the polar and equatorial deformations, respectively, 
\begin{equation}\label{j2r}
J_2^{r} = k_f \frac{\Omega^2R^3}{3Gm_1}
\end{equation}
is the rotational deformation, and
\begin{eqnarray}\label{max2}
&&J_2^t=k_f \frac{m_0}{2m_1}\left(\frac{R}{r_1}\right)^3 \ ,\\
&&C_{22}^t=\frac{k_f}{4}\frac{m_0}{m_1}\left(\frac{R}{r_1}\right)^3\cos2\gamma \ ,\\
&&S_{22}^t=-\frac{k_f}{4}\frac{m_0}{m_1}\left(\frac{R}{r_1}\right)^3\sin2\gamma \ ,
\end{eqnarray}
are the tidal equilibrium values for the gravity coefficients \citep{Correia_Rodriguez_2013},
where $k_f$ is the fluid second Love number.

\section{Aplication to the CoRoT-7 system}
\label{corot}

We apply the model from previous section to the CoRoT-7 planetary system, which is composed by two short-period planets. 
CoRoT-7 is a young G9V sun-like star with mass $m_0=0.915 \pm 0.019\,M_{\odot}$, radius $R_0=0.818 \pm 0.016 \, R_{\odot}$, and age of $1.32 \pm 0.76$~Gyr \citep{Barros_etal_2014}.

\subsection{Observed system}

\begin{table}
\begin{center}
\caption{\small The adopted current orbital elements and physical data of the CoRoT-7 system
\citep{Barros_etal_2014, Haywood_etal_2014}.
}\label{tabela-corot}
\begin{tabular}{c c c c c}
\hline
   Body & $m_i$  & $R$ & $a_{i\,{\textrm{\scriptsize current}}}$ (AU)& $e_{i\,{\textrm{\scriptsize current}}}$\\
  \hline
  0 & $0.915 \pm 0.019\,M_{\odot}$  & $0.82\,R_{\odot}$ & - & -\\
 
  1 & $4.73 \pm 0.95\,M_{\oplus}$  & $1.53\,R_{\oplus}$ & 0.0171 & $0.12\pm0.07$ \\
 
  2 & $13.56 \pm1.08\,M_{\oplus}$  & - & 0.0455 & $0.12\pm0.06$ \\
  \hline
\end{tabular}
\end{center}
\end{table}

The system was observed combining radial velocity and transit measurements \citep{Barros_etal_2014, Haywood_etal_2014}, which provide us the radius and the true mass of the inner planet, hence an estimation of its density. 
The inner planet, CoRoT-7\,b, and the outer planet, CoRoT-7\,c, have masses $m_1=4.73\,M_{\oplus}$ and $m_2=13.56\,M_{\oplus}$, respectively \citep{Haywood_etal_2014}, whereas the radius of CoRoT-7\,b is $R=1.53\,R_{\oplus}$ \citep{Barros_etal_2014}. 
Within the uncertainties of the observations, the mean density of the inner planet is $6.6 \pm 1.5$~g/cm$^3$ \citep{Haywood_etal_2014}, i.e., equal or larger than the density of the Earth.
We can thus assume that CoRoT-7\,b is a rocky planet in the super-Earth mass regime. 

The orbital periods of the planets are $P_{1\,\textrm{\scriptsize orb}}=20.5$~h and $P_{2\,\textrm{\scriptsize orb}}=3.70$~d \citep{Haywood_etal_2014}.
The best fit to the observational data determines that both planets evolve in non-circular orbits with an eccentricity value around 0.1 \citep{Haywood_etal_2014}, although the error bars are large and these values are still compatible with zero.
Since both planets are very close to the star, the usual expectation is that the orbits become circular after some time \citep{Ferraz-Mello_etal_2011}.
However, CoRoT-7 is a young star, and some transient equilibria for the eccentricity can occur, which could explain the non circular orbits at present.

Although the inclination of the CoRoT-7\,c planet is not yet determined, for simplicity we assume that the orbits of the planets are coplanar.
All adopted physical and orbital parameters for the system are listed in Table \ref{tabela-corot}. 



\subsection{Numerical simulations}

We performed a series of numerical simulations using the set of equations (\ref{mov1}) to (\ref{max1}).
As in previous studies \citep[e.g.,][]{Ferraz-Mello_etal_2011, Rodriguez_etal_2011, Dong_Ji_2012}, the idea is to study the past evolution of the CoRoT-7 system and figure out how the orbits evolved into the present ones.
Since the initial system configuration is unknown, we take different initial values for the orbital parameters. 

In the following, we denote the semi-major axis and the eccentricity by $a$ and $e$, respectively.
For the initial semi-major axes, we assume $a_1=0.0188$ AU and $a_2=0.0455$ AU. 
Since the orbital angular momentum of the system, $L$, is conserved along the evolution 
(the rotational angular moment can be neglected in comparison), 
we have $L = L_1 + L_2$, where
\begin{equation}
\vec{L}_i \approx L_i \, \vec{k} = m_i \sqrt{\mu_i a_i (1-e_i^2)} \, \vec{k}
\ . \label{160324z}
\end{equation}
The eccentricity of the outer planet can then be obtained through the current elements listed in Table \ref{tabela-corot} as
\begin{equation}\label{am}
e_2\simeq\left[1-\left(\frac{L}{m_2\sqrt{\mu_2 a_2}}-\frac{m_1}{m_2}\sqrt{\frac{\mu_1 a_1}{\mu_2 a_2}(1-e_1^2)}\right)^2 \right]^{1/2} \ ,
\end{equation}
where $L$ is computed from the present values of the orbital elements (Table\,\ref{tabela-corot}).

We start with an initial rotation rate such that $\Omega/n_1=4.1$, where $n_1$ is the orbital mean motion of the inner planet.
This value for the rotation is not critical, as the spin quickly evolves under tides into a SOR.
Other adopted initial values (also not critical) for the numerical simulations are\footnote{\bfx{Due to the computational cost of the numerical simulations, we are not able to explore all the unknown parameters.
However, we performed some runs with changes in these parameters, without observing any relevant changes in the evolution.}}: $\theta=0^\circ$, $v_1=v_2=0^\circ$, $\varpi_1=10^{\circ}$, $\varpi_2=100^{\circ}$, $k_f=1.0$ and $C=\xi m_1 R^2$, with $\xi=0.35$.
In order to overcome our total ignorance on the values of the relaxation times, we perform numerical simulations with six values of $\tau=10^{-3}$, $10^{-2}$, $10^{-1}$, $10^{0}$, $10^{1}$, and $10^{2}$\,yr.

Since CoRoT-7\,b is a super-Earth, we also assume non-zero values for the permanent non-spherical figure of the planet.
For Venus (which rotates slowly and thus we can neglect the effect of the rotation on its shape) we have $J_2^0 \sim C_{22}^0 \sim 10^{-6}$ \citep{Yoder_1995cnt}.
However, for a more massive super-Earth we expect these values to become even smaller due to a stronger gravity at the surface.
We thus adopt here $J_2^0=C_{22}^0=10^{-7}$.
These values correspond almost to a quasi-spherical shape for the unperturbed planet, but they still facilitate the capture in SORs.

\subsubsection{High initial $e_1$}
\label{hie}

\begin{figure*}
\begin{center}
\begin{tabular}{cc}
\includegraphics[height=\columnwidth,angle=270]{\figpath 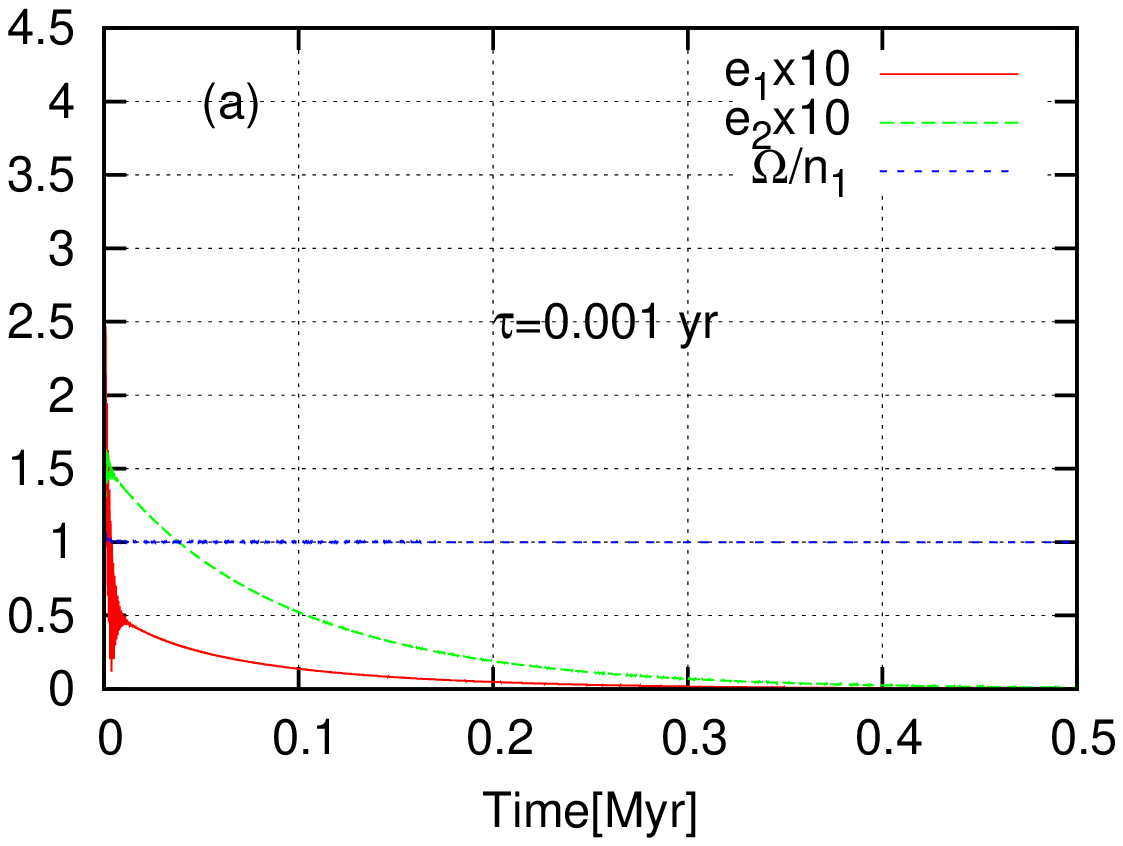} &
\includegraphics[height=\columnwidth,angle=270]{\figpath 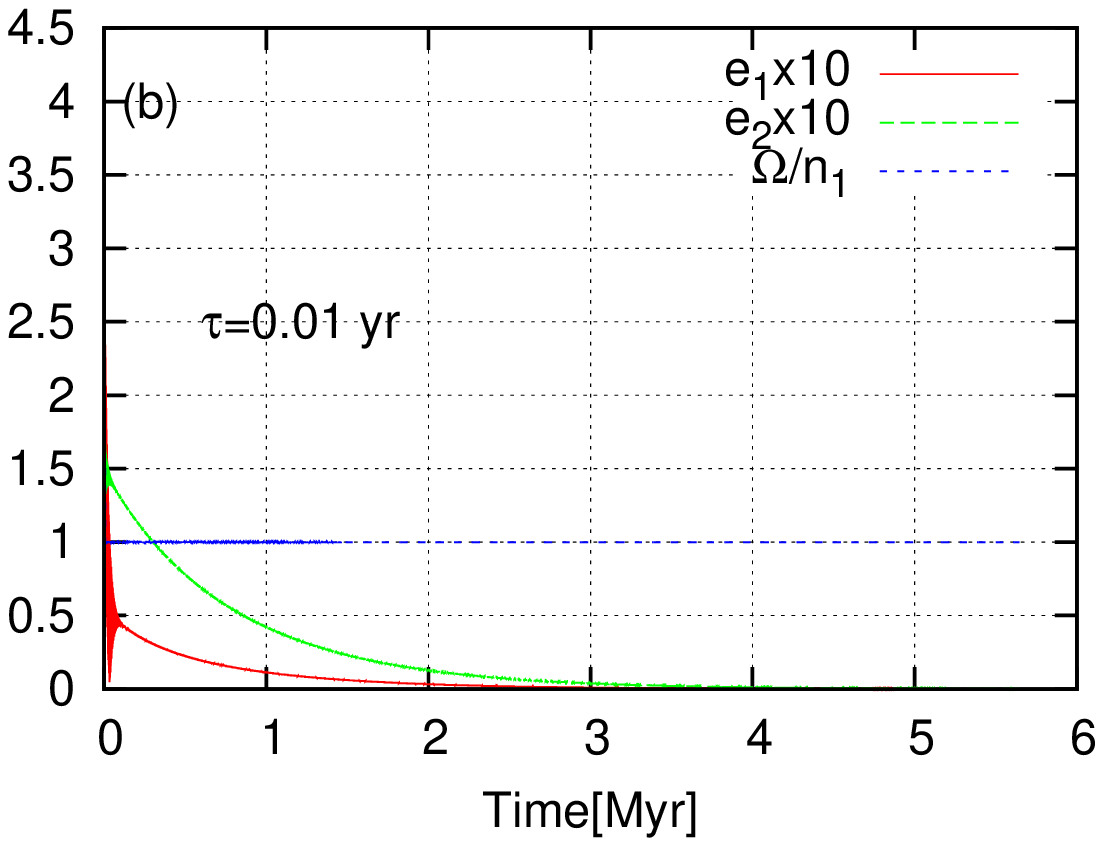} \\
\includegraphics[height=\columnwidth,angle=270]{\figpath 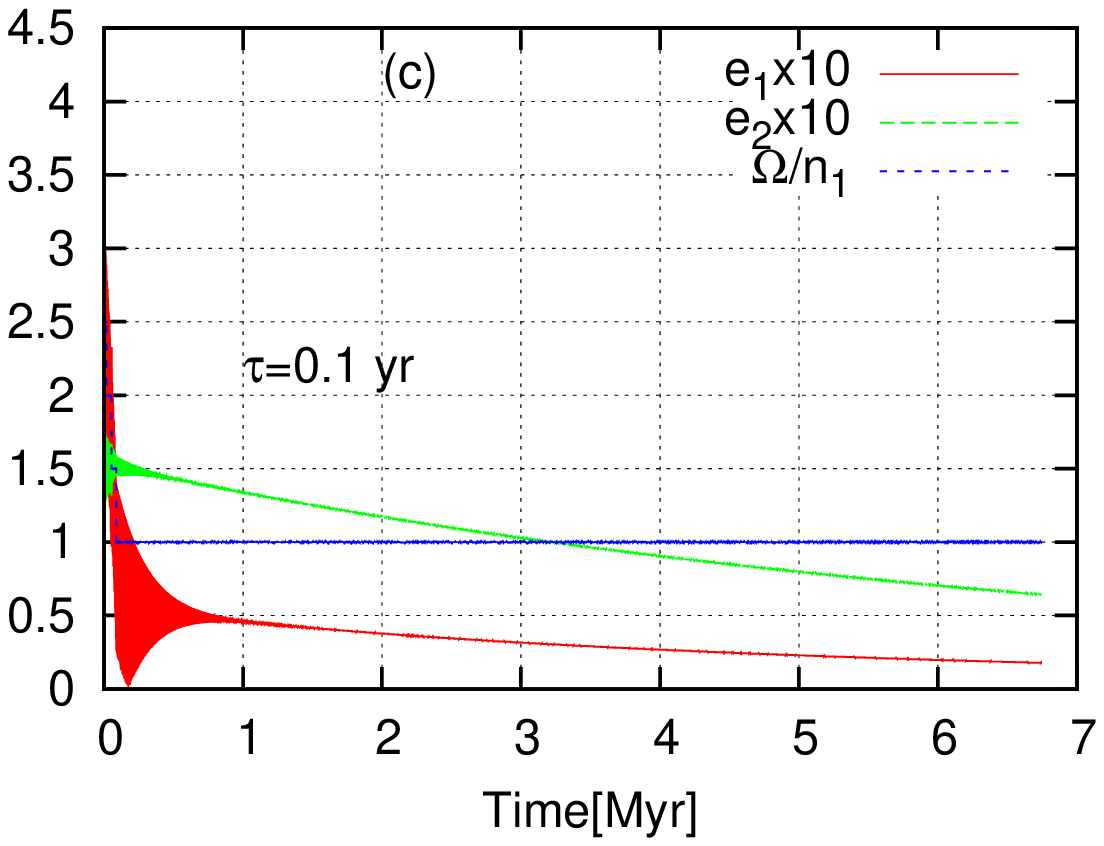} &
\includegraphics[height=\columnwidth,angle=270]{\figpath 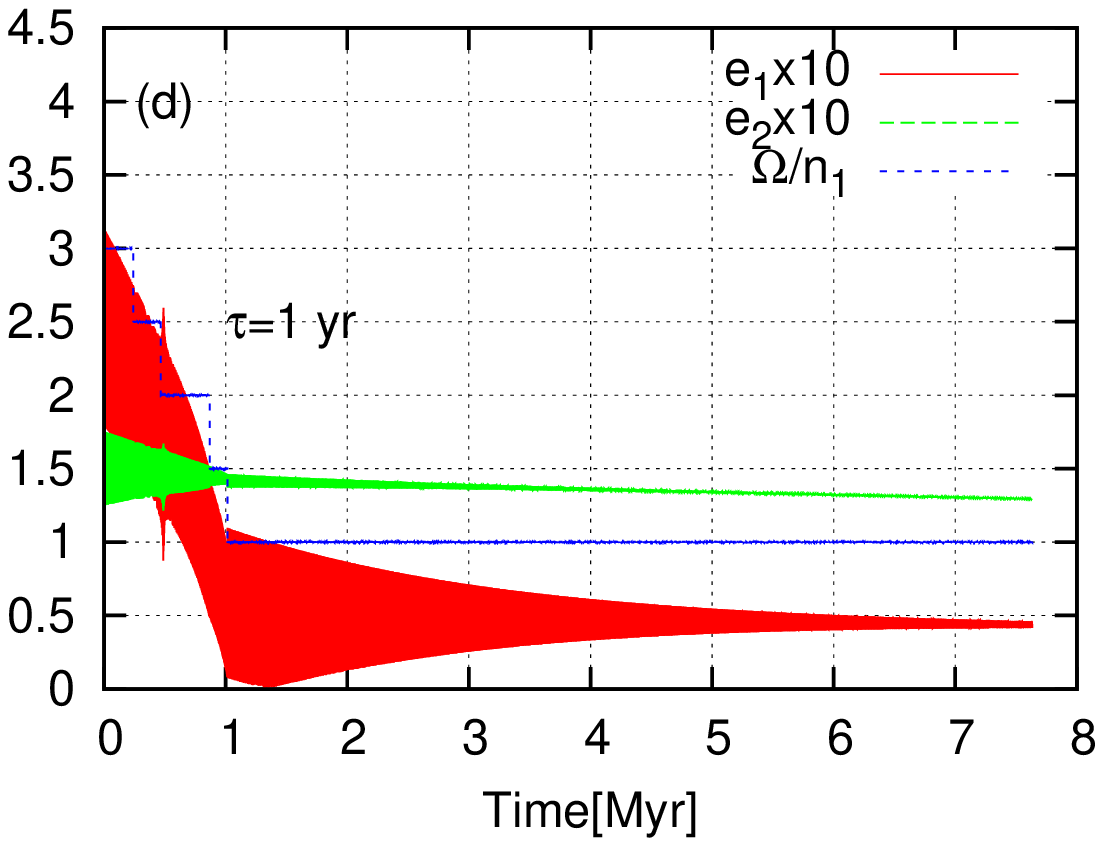} \\
\includegraphics[height=\columnwidth,angle=270]{\figpath 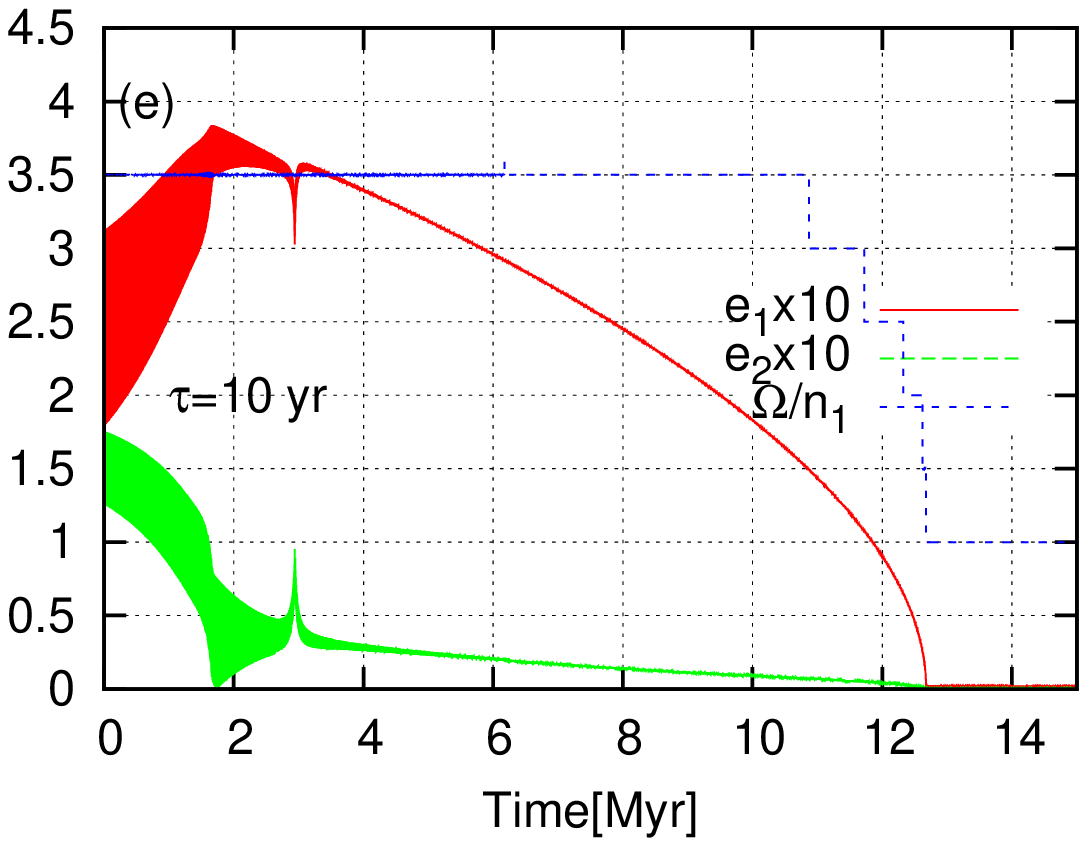} &
\includegraphics[height=\columnwidth,angle=270]{\figpath 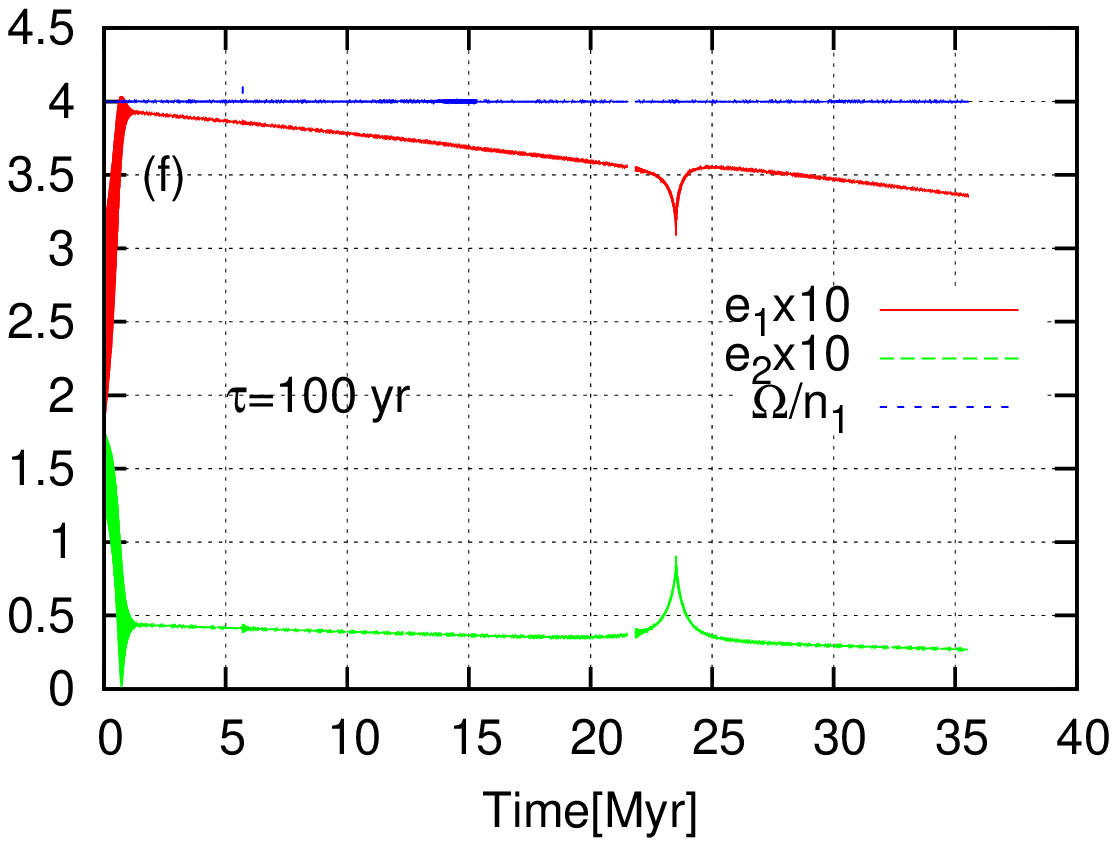}
\end{tabular}
\caption{\small Evolution of the eccentricities and rotation rate with time for 
six values of the relaxation time $\tau$. 
For small $\tau$ (panels (a) and (b)), the orbits are quickly circularized and the rotation rate trapped in the synchronous motion.
For intermediate $\tau$ (panels (c) and (d)) the rotation is temporarily captured in SORs, which are destabilized as the eccentricity decays, ending with synchronous motion.
For large $\tau$ (panels (e) and (f)), the eccentricity of the inner orbit is excited to high values. 
In panel (e) the eccentricities are perturbed near 3~Myr because the system crosses the 4:1 mean-motion resonance.}
\label{ecce-rot}
\end{center}
\end{figure*}

The CoRoT-7 planetary system most likely formed away from the star and then migrated inward \citep[e.g.][]{Terquem_Papaloizou_2007, McNeil_Nelson_2010, Cossou_etal_2014}.
In this process, the planets can be trapped in mean motion resonances, which increase the eccentricities until the resonance is broken \citep{Beauge_etal_2003, Ferraz-Mello_etal_2003}.
Therefore, we first consider the case of initial high eccentricity for the inner planet. 
For $e_1=0.25$, we obtain from expression (\ref{am}), $e_2=0.1546$.
In Fig.~\ref{ecce-rot} we plot the temporal evolution of the eccentricities and the ratio $\Omega/n_1$ for all values of $\tau$ ($10^{-3}-10^2$\,yr).

In panel (a), corresponding to $\tau = 10^{-3}$\,yr, the planet is in the low-frequency regime since $n_1 \tau < 1$.
In this regime, the orbital evolution of the system is expected to be similar to the linear tidal model, for which the tidal dissipation is proportional to the corresponding tidal frequency \citep[e.g.,][]{Singer_1968, Mignard_1979}.
According to this model, the rotation of the planet evolves into an equilibrium value that depends on the eccentricity of the orbit, often called the pseudo-synchronization, for which \citep[e.g.,][]{Correia_etal_2011} 
\begin{equation}
\frac{\Omega}{n_1} = \frac{1 + \frac{15}{2}e_1^2 + \frac{45}{8}e_1^4 + \frac{5}{16}e_1^6}{(1 + 3e_1^2 + \frac38e_1^4)(1-e_1^2)^{3/2}} = 1+6e_1^2 + {\cal O} (e_1^4) \ . \label{090520a}
\end{equation}
Our simulations confirm that the rotation of the planet follows this equilibrium, which is always faster than the synchrounous rotation unless the orbit becomes circular.
However, since we are considering a residual value for the $C_{22}$, each time the rotation crosses a SOR there is a chance of capture, although very small because $C_{22} = 10^{-7}$ \citep[see][]{Goldreich_Peale_1966, Rodriguez_etal_2012}.

For the Earth and Mars, we have $\tau\sim10^{-1}$\,yr \citep{Correia_etal_2014}.
Moreover, although $\tau\sim10^{-1}$\,yr provides a good estimation for the average present dissipation ratios on these two planets, it appears to be incoherent with the observed deformation.
Indeed, in the case of the Earth, the surface post-glacial rebound due to the last glaciation about $10^4$~years ago is still going on, suggesting that the Earth's mantle relaxation time is something like $\tau \approx 4,400$\,yr \citep{Turcotte_Schubert_2002}.
For rocky planets a value of $\tau=10^{-3}$\,yr is thus very unlikely, and it is better to consider higher values for $\tau$.
For all the remaining adopted values, we have that $n_1 \tau  >1$, that is, the planet is in the high-frequency regime.
In this regime, the tidal energy dissipated is inversely proportional to the frequency.

In panels (b) and (c), corresponding to $\tau = 10^{-2}$\,yr and $10^{-1}$\,yr, respectively, we still observe a \bfx{rapid} synchronization of the rotation with the orbital motion ($\Omega/n_1=1$), 
while both eccentricities are quickly damped to zero (orbital circularization). 
The only difference is that in panel (c) the rotation becomes captured 
in higher order SORs ($\Omega/n_1=$ 5:2, 2:1, 3:2) at the beginning of the simulation, that are nevertheless quickly destabilised until the spin reaches the synchronization.
Dissipation of the tidal energy only occurs in the inner planet, but both eccentricities are damped since the system is coupled.
In the beginning of the simulations, $e_1$ is damped more efficiently, but when it approaches zero, the pericenters of the planets become aligned, and both eccentricities approach a quasi-equilibrium value.
The transfer of angular momentum between the two orbits becomes more efficient and both eccentricities are damped together \citep[for more details see][]{Mardling_2007,Laskar_etal_2012}.
The difference in the orbital time-scales is accounted for the value of the relaxation time adopted in each simulation.

In panel (d), corresponding to $\tau = 1$\,yr, we observe that the rotation evolves through a succession of temporary trappings in SORs (3:1, 5:2, 2:1, 3:2), ending with synchronous motion (1:1). 
In this case, the rotation spends more time trapped in higher order resonance than for $\tau=10^{-1}$\,yr. All the resonances are destabilized as the eccentricity decays, in agreement with
previous results \citep{Rodriguez_etal_2012, Correia_etal_2014}, because the capture and escape probability in SORs critically depends on the eccentricity \citep[e.g.,][]{Goldreich_Peale_1966, Correia_Laskar_2009}.

In panels (e) and (f), corresponding to the largest values of $\tau$, we observe that the rotation is captured in high order SOR.
For $\tau=10$ yr, the rotation is initially trapped in the 7:2 SOR and for $\tau=10^2$ yr it is initially trapped in the 4:1 SOR.
As explained in \citet{Correia_etal_2014}, large $\tau$ imply that
the relaxation time is much longer than the orbital period, allowing the prolateness of the planet to acquire a much larger deformation than the permanent $C_{22}^0 = 10^{-7}$ value.
This helps the rotation to be captured more easily in SOR. 
The rotation is also trapped for longer periods of time because SOR are only destabilized for very low eccentricity values \citep[see][]{Correia_etal_2014}.

Unlike previous simulations for lower $\tau$ values, in panels (e) and (f) we also observe that the eccentricity of the inner orbit is initially excited to a high value, whereas the outer planet eccentricity is simultaneously damped (due to the angular momentum conservation). 
The initial excitation of $e_1$, that we call ``eccentricity pumping", is \bfx{somewhat} unexpected, since most studies on tidal evolution of the orbits predict that the eccentricities can only be damped \citep[e.g.,][]{Kaula_1964, Mignard_1979, Hut_1981}.
When the outer orbit eccentricity approaches zero, the pericenters of the planets become anti-aligned, the eccentricity pumping ceases, and both eccentricities are slowly damped to zero as in the previous cases.

A similar initial excitation for the eccentricity has already been reported for gaseous planets within $0.1 < a_1 <0.3$~AU \citep{Correia_etal_2012, Correia_etal_2013, Greenberg_etal_2013}.
For these kind of planets, the linear tidal model is well suited. As a consequence, the eccentricity pumping \bfx{is related to a variation in the $J_2$} of the inner planet due to the rotational deformation (Eq.\,(\ref{j2r})), that tends to follow the pseudo-synchronous equilibrium (Eq.\,(\ref{090520a})). 
However, in the present case the rotation is locked in a SOR, so this effect can be neglected.
Moreover, here the pumping effect appears in a system with a very close-in super-Earth and assuming a viscoelastic response. 
In section~\ref{pumping} we explain this eccentricity pumping in detail, and show that this effect \bfx{is still related to a variation in $J_2$}, but as a result of the tidal deformation term (Eq.\,(\ref{max2})).

\begin{figure}
\begin{center}
\includegraphics[height=\columnwidth,angle=270]{\figpath 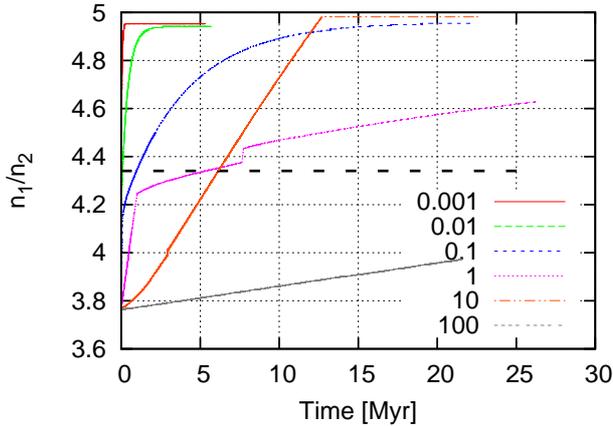}
\caption{\small Evolution of the ratio of mean orbital motions with time for all adopted values of $\tau$.
\bfx{The labeled colors correspond to the set of adopted values of $\tau$ in units of yr.}
Larger values of $\tau$ delay the evolution of the system.
In addition, once the rotation is trapped is the synchronous motion the evolution slows down even more.
The dashed line gives the present observed value (Table~\ref{tabela-corot}).}
\label{n1n2}
\end{center}
\end{figure}

Because the process of tidal circularization is slower for large values of $\tau$ (Fig.\,\ref{n1n2}), \bfx{in panel (f)} we are not able to show the complete evolution of the rotation. 
However, we expect that the process of synchronization follows a similar behavior as in the previous panels,
following its evolution under subsequent lower order captures (3:1, 5:2, 2:1, 3:2) to finally reach the 1:1 SOR.
\bfx{We note that, despite the capture into high-order SOR, the eccentricity pumping is not related with such trappings (see Sect. \ref{pumping}).} 

In Fig.~\ref{n1n2}, we plot the temporal evolution of the ratio of mean orbital motions, $n_1/n_2$.
This figure allows us to better compare the orbital evolution time-scales for each value of $\tau$, and also to see the impact of the orbital 4:1 mean motion resonance crossing.
The dashed line in Fig.~\ref{n1n2} gives the present observed value (Table~\ref{tabela-corot}).
In general, the orbital decay is faster for small values of $\tau$, since the dissipation is inversely proportional to $\tau$ in the high frequency regime ($\tau n_1 > 1$).
However, it is interesting to note that in all simulations there is a regime transition for which the evolution of the ratio $n_2/n_1$ slows down.
This corresponds to the moment at which the rotation is captured in the synchronous resonance, since dissipation of tidal energy only occurs on the orbit \citep[see][]{Rodriguez_etal_2012}.
\bfx{For large $\tau$ values, this transition only occurs when the orbit is nearly circularized, since higher-order resonances are stable for very low ecentricity values \citep{Correia_etal_2014}.
As a consequence, the orbital decay occurs faster for $\tau = 10$\,yr than for $\tau = 1$\,yr.}
Nevertheless, for $\tau \ge 100$\,yr, the system takes a long time to attain the present configuration, which may explain why the eccentricity of the inner orbit is not yet fully damped.

\begin{figure*}
\begin{center}
\includegraphics[height=\columnwidth,angle=270]{\figpath 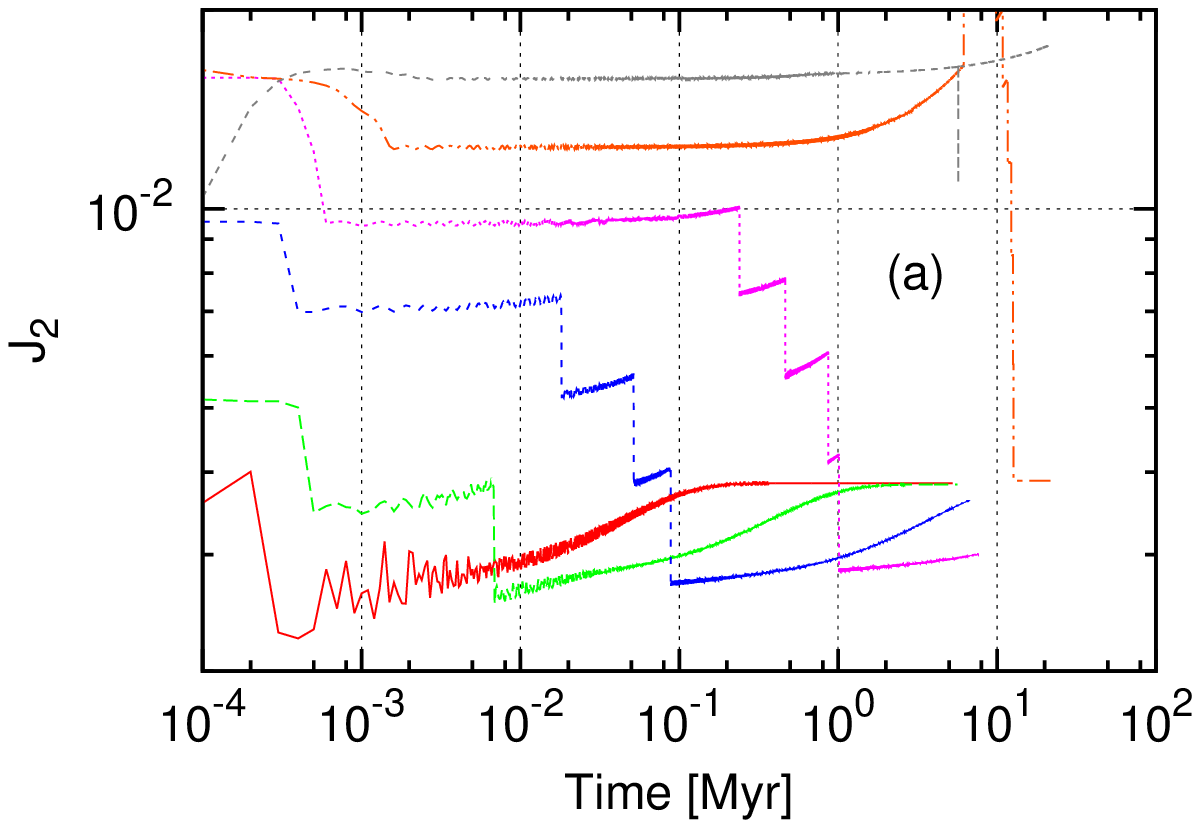}
\includegraphics[height=\columnwidth,angle=270]{\figpath 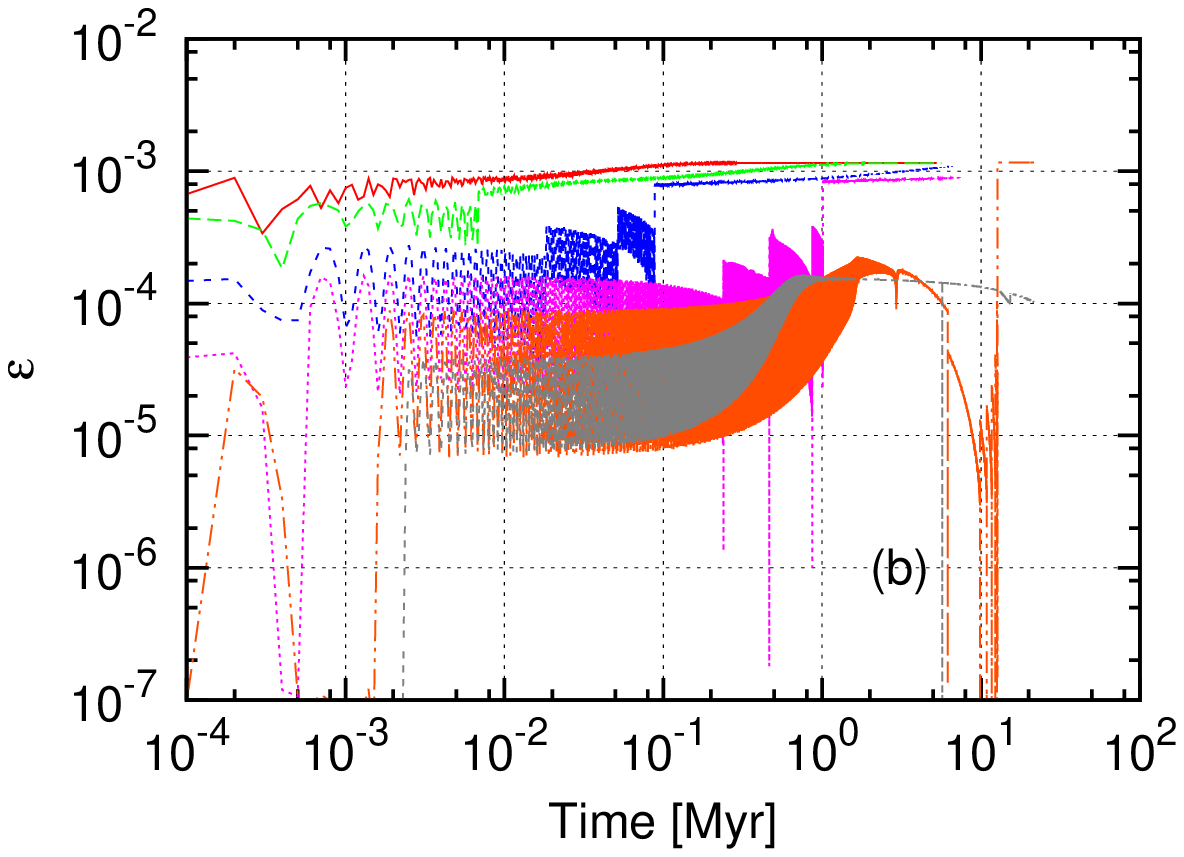}
\includegraphics[height=\columnwidth,angle=270]{\figpath 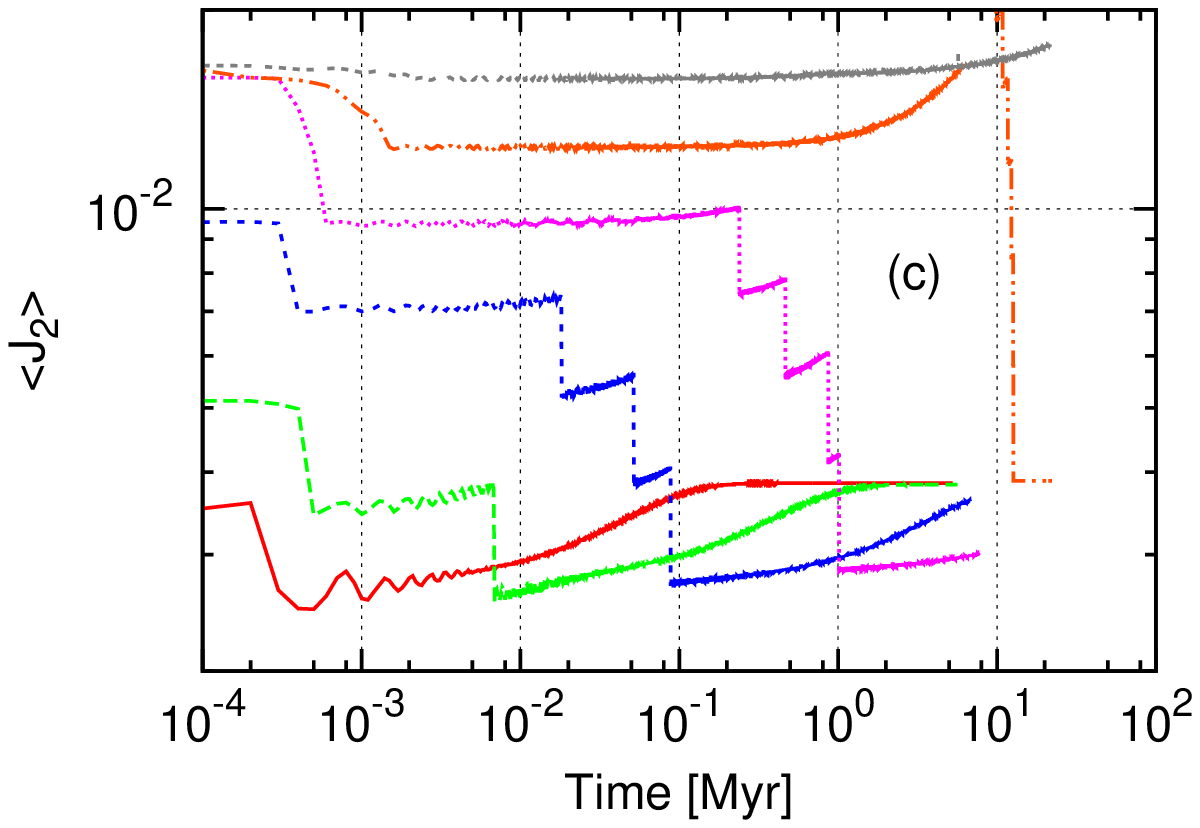}
\includegraphics[height=\columnwidth,angle=270]{\figpath 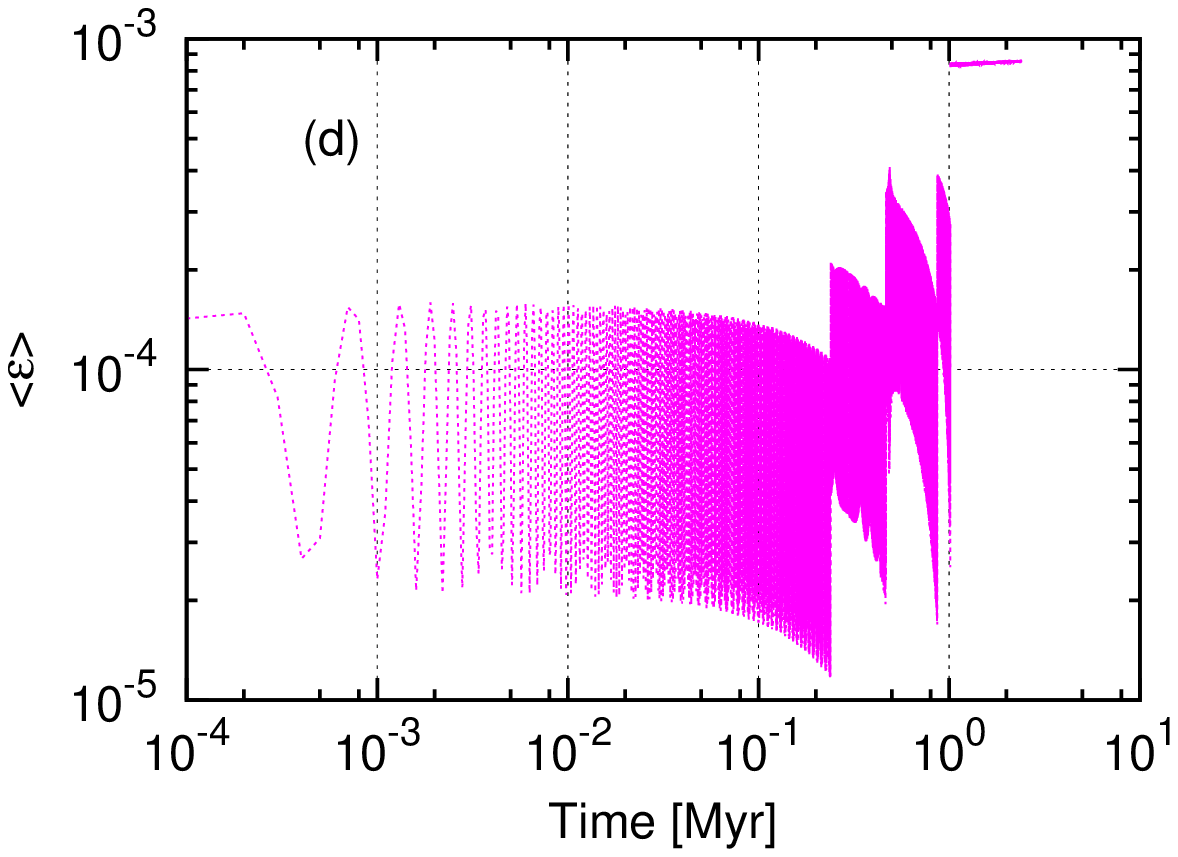}
\caption{\small Evolution of the shape of the planet with time for all adopted values of $\tau$. 
We show the instantaneous values of $J_2$ (a) and $\epsilon$ (b), together with their average values $\langle J_2\rangle$ (c) and $\langle \epsilon \rangle$ (d) (Eqs.\,(\ref{j2}) and (\ref{c22})).
The colors correspond to the same values of $\tau$ as in Fig.~\ref{n1n2}. 
For simplicity, in the case of $\langle\epsilon\rangle$ we only show the results for $\tau=1$\,yr in order to better visualize the changes in the shape under the sequence of resonant trappings of the rotation. 
The instantaneous values of $J_2$ and $\epsilon$ are in excellent agreement with the theoretical prediction for the averaged values.}
\label{eps}
\end{center}
\end{figure*}

In Fig.~\ref{eps}, we plot the evolution of the instantaneous shape of CoRoT-7\,b as a function of time, given by its oblateness, $J_2$, and prolateness,
\begin{equation}
\epsilon=\sqrt{C_{22}^2+S_{22}^2}
\ . \label{160520a} 
\end{equation}
To better understand the different behaviors, we also plot the average of the equilibrium shape over one orbital period \citep[see][]{Correia_etal_2014}.
For $J_2$ we have
\begin{equation}\label{j2}
\langle J_2 \rangle = J_2^0 + J_2^r + \langle J_2^t \rangle  \ , 
\end{equation}
where
\begin{equation}\label{j2e}
\langle J_2^t \rangle = {\cal A} \, (1-e_1^2)^{-3/2} \ ,
\end{equation}
and
\begin{equation}
{\cal A} = \frac{k_f}{2} \frac{m_0}{m_1}\left(\frac{R}{a_1}\right)^3
\ . \label{160328d}
\end{equation}
The mean equilibrium value of $\epsilon$ depends on the SOR in which the rotation 
is trapped in. 
With $p=\Omega/n_1$ we have
\begin{equation}\label{c22}
\langle \epsilon \rangle = C_{22}^0 + \frac{\cal A}{2}  X_{2p}^{-3,2}(e_1) \ ,
\end{equation}
where $X_k^{l,m} (e_1) $ are Hansen coefficients such that
\begin{equation}
\left( \frac{r_1}{a_1} \right)^{l} \mathrm{e}^{\mathrm{i} m \nu_1} =  \sum^{+\infty}_{k=-\infty} X_k^{l,m} (e_1) \, \mathrm{e}^{\mathrm{i} k M_1} 
\ . \label{061120gb}
\end{equation}

The $J_2$ and $\epsilon$ obtained numerically and analytically (Eqs.\,(\ref{j2}) and (\ref{c22})) show that the instantaneous values closely follow their average equilibrium values (Fig.\,\ref{eps}). 
The sudden variations observed correspond to the transition between two successive SORs.
When the rotation jumps from a SOR to a lower order one, the $J_2$ decreases, which is a consequence of the term $J_2^r$ (Eq.\,(\ref{j2r})), that is proportional to $(\Omega/n_1)^2=p^2$.
On the other hand, for a given SOR, the $J_2$ increases as the inner planet migrates towards the central star. 
From expression (\ref{j2e}) we see that $\langle J_2^t \rangle$ must increase as $a_1$ decreases, despite the influence of the factor depending on $e_1$. 
At the end of the evolution, when the rotation becomes synchronous, the planet acquires the same $J_2$ value in all scenarios, because whatever the value of $\tau$ is, the planet has enough time to reach the equilibrium figure.

Unlike the $J_2$ variations, the prolateness ($\epsilon$) increases when the rotation changes from one SOR to the next (lower order) one. 
The prolateness of the planet also follows the average equilibrium value for each SOR (Eq.\,(\ref{c22})).
It temporarily decreases with the eccentricity, since $X_{2p}^{-3,2} (e)$ is a decreasing function with $e$.
However, when the critical eccentricity for each resonance is attained, 
$\epsilon$ increases again because $X_{2p}^{-3,2} (e) \propto e^{2(p-1)}$ \citep[e.g.][]{Correia_etal_2014}.
When the synchronous rotation is reached, the deformation always points along the direction of the star and it grows a lot, since $X_{2}^{-3,2} (e) \approx 1 - 5e^2/2$, i.e., the prolateness marginally depends on the eccentricity.
Again, it becomes the same for all $\tau$ values, since the planet has enough time to reach the maximal deformation.

\subsubsection{Low initial $e_1$}
\label{sie}

We now suppose that the initial eccentricity of the inner orbit is low. 
This assumption can be justified, among other reasons, considering a scenario where the orbit of the inner planet was not excited by any mean motion resonance with the outer planet, and therefore it kept a low eccentricity value during the migration process.
We thus take a low initial value of $e_1=0.05$ and keep the same previous values of initial semi-major axes. 
Applying Eq. (\ref{am}), we obtain \bfx{the initial value} $e_2=0.1933$.

\begin{figure}
\begin{center}
\includegraphics[height=\columnwidth,angle=270]{\figpath 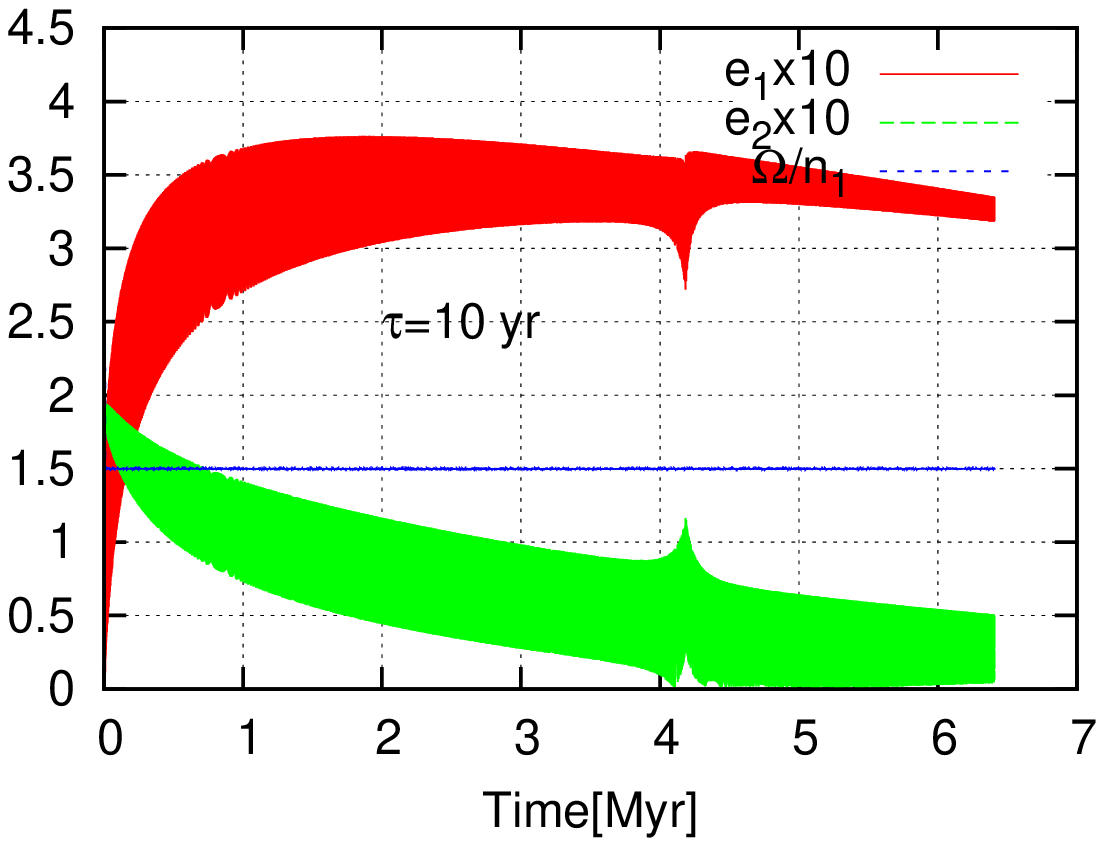}
\includegraphics[height=\columnwidth,angle=270]{\figpath 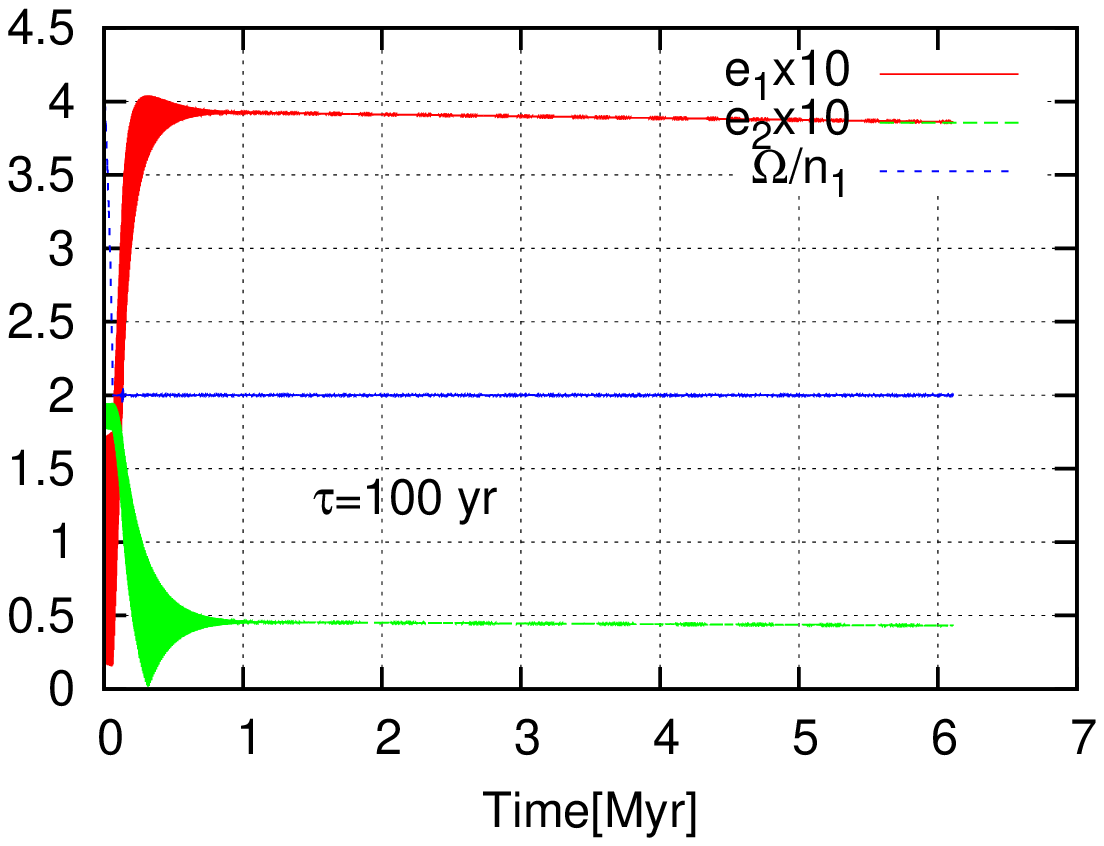}
\includegraphics[height=\columnwidth,angle=270]{\figpath 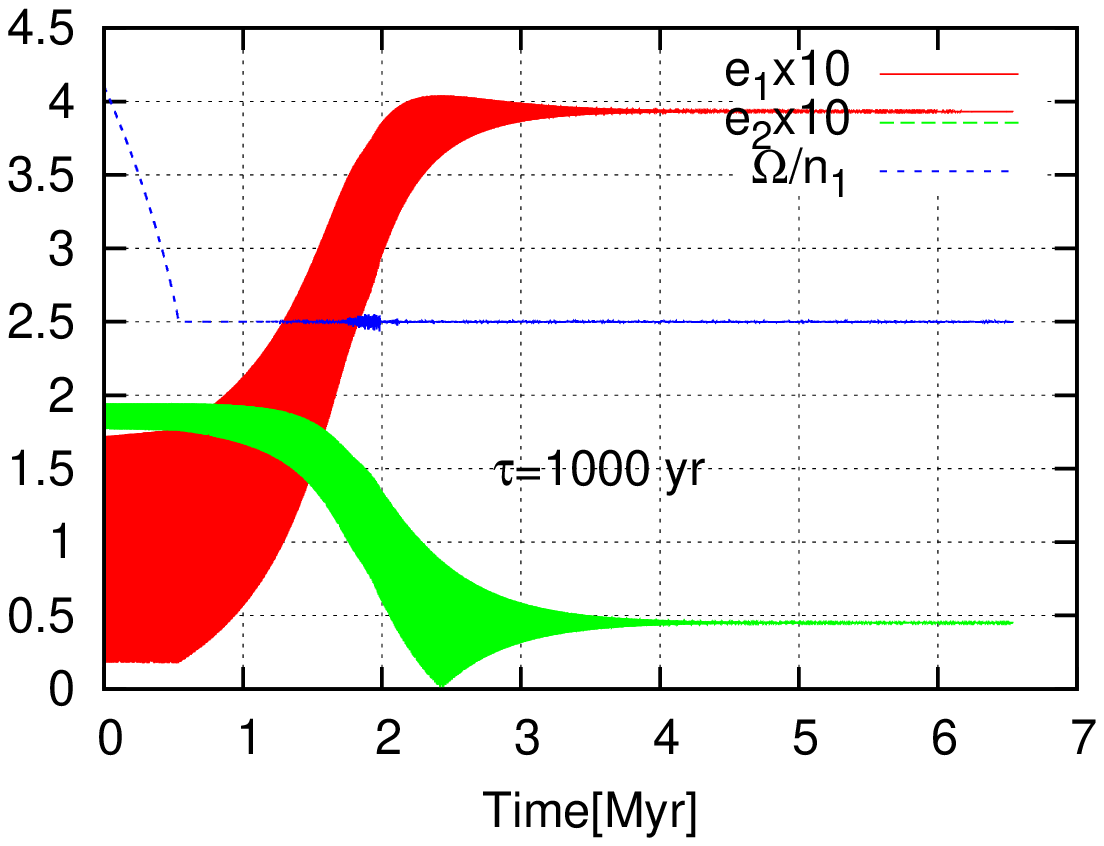}
\caption{\small Evolution of the eccentricities and rotation rate for $\tau=10$\,yr,  $10^2$\,yr and $10^3$\,yr, and low initial eccentricity ($e_1=0.05$). 
The eccentricity pumping also appears in this case, indicating that it may be an efficient mechanism that took place during the evolution of the CoRoT-7 planetary system.}
\label{ecce-rot-low}
\end{center}
\end{figure}

For $\tau < 1$\,yr, the eccentricity can only be damped (Fig.\,\ref{ecce-rot}), so these cases are not interesting to study here again.
However, for $\tau \ge10$\,yr, we observed a strong increase in $e_1$.
In order to check if the initial eccentricity pumping is still present for low initial eccentricity,
in Fig.~\ref{ecce-rot-low} we plot the evolution of eccentricities and rotation for $\tau = 10$\,yr to $10^3$\,yr.
The initial increase in $e_1$ is still observed in all cases, so we conclude that the pumping effect is an efficient mechanism that \bfx{may have occurred} during the past evolution of the CoRoT-7 
system\footnote{We also performed a simulation (not shown here) with $a_1=0.02$ AU and the eccentricity pumping also appeared.}.
Since the eccentricity pumping is present even for initial low eccentricities of the inner orbit, \bfx{it provides a possible explanation} for the present observed high value of 0.12 (Tab.\,\ref{tabela-corot}). 
It can also explain the occurrence of high eccentricities of other close-in super-Earths in multiple planetary systems.

The main difference with respect to the case with high initial eccentricity (section~\ref{hie}) is that the rotation is now initially captured in lower order SORs.
This behavior was expected due to the lower initial value of $e_1$, for which high-order SORs are unstable. 
Once captured in a SOR, the pumping effect helps to keep the rotation trapped in a non-synchronous spin-orbit resonant configuration for a longer period of time as well.

\begin{figure*}
\begin{center}
\includegraphics[height=\columnwidth,angle=270]{\figpath 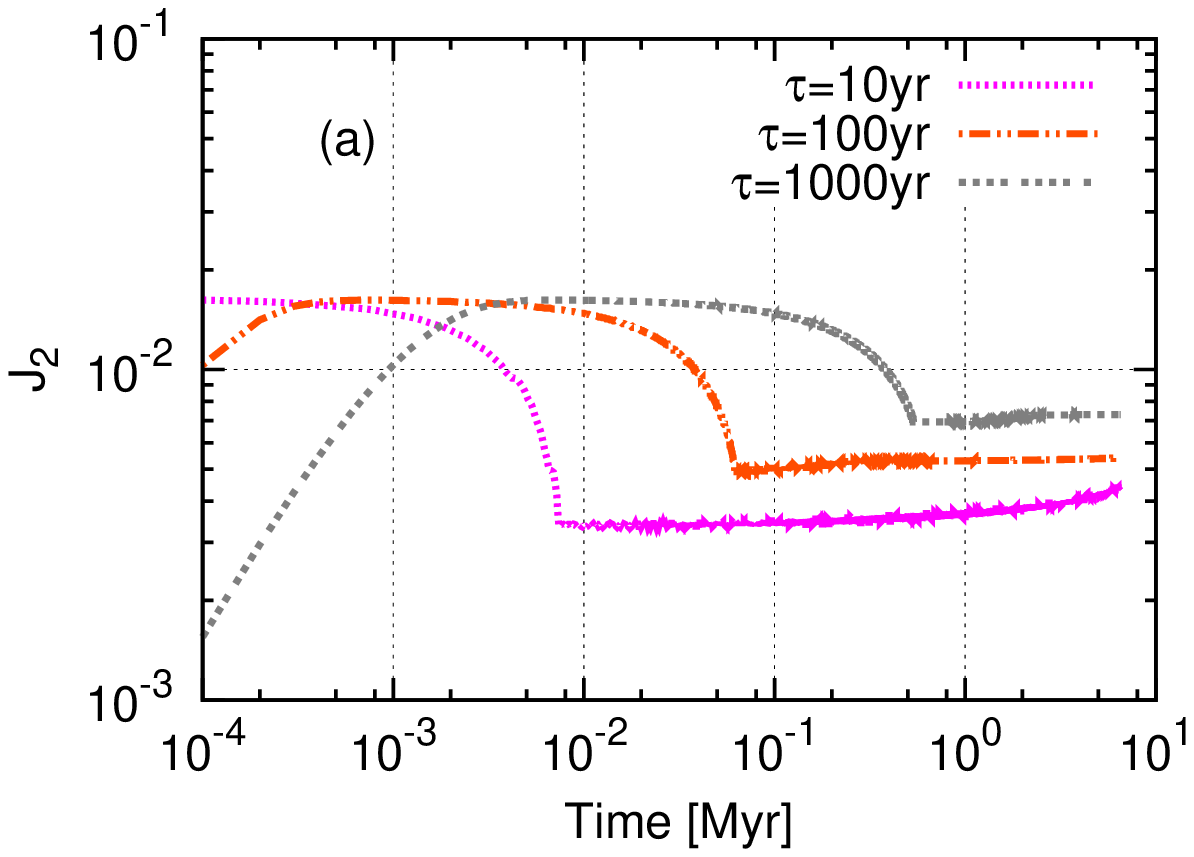}
\includegraphics[height=\columnwidth,angle=270]{\figpath 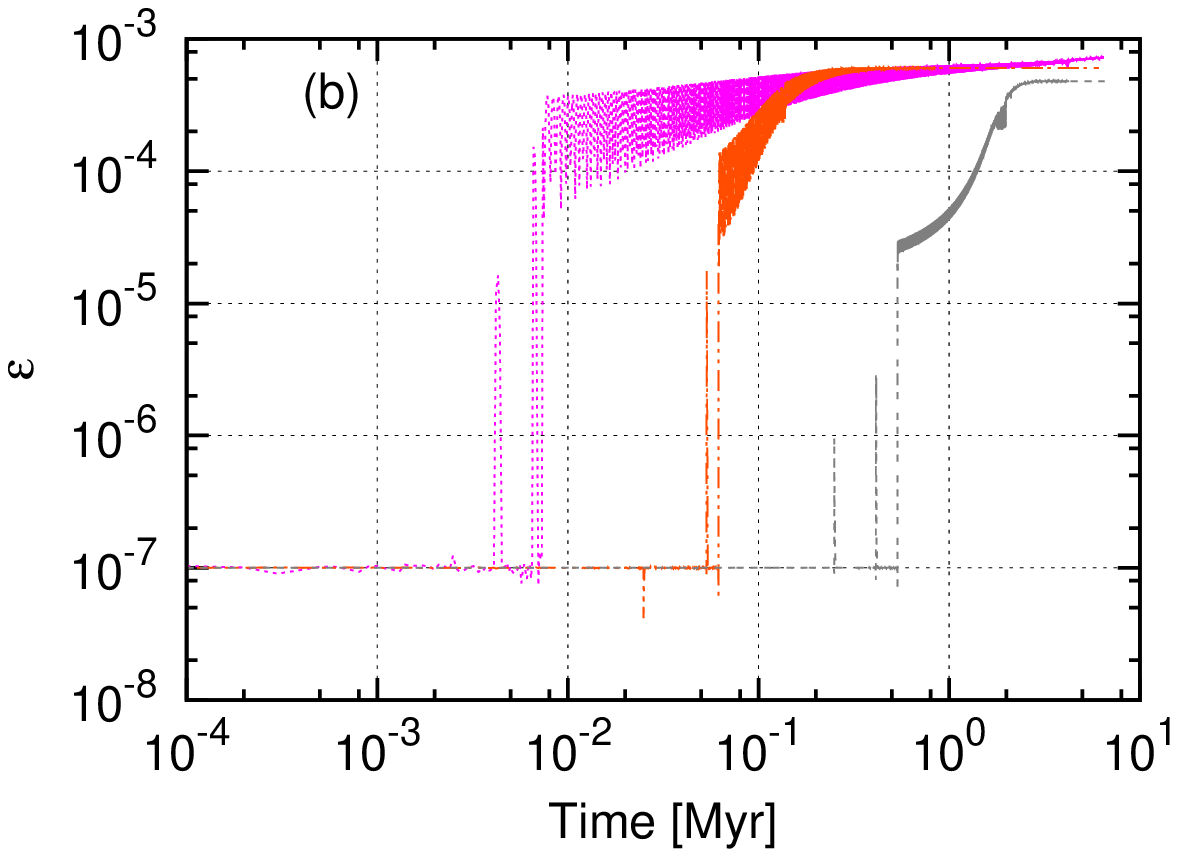}
\includegraphics[height=\columnwidth,angle=270]{\figpath 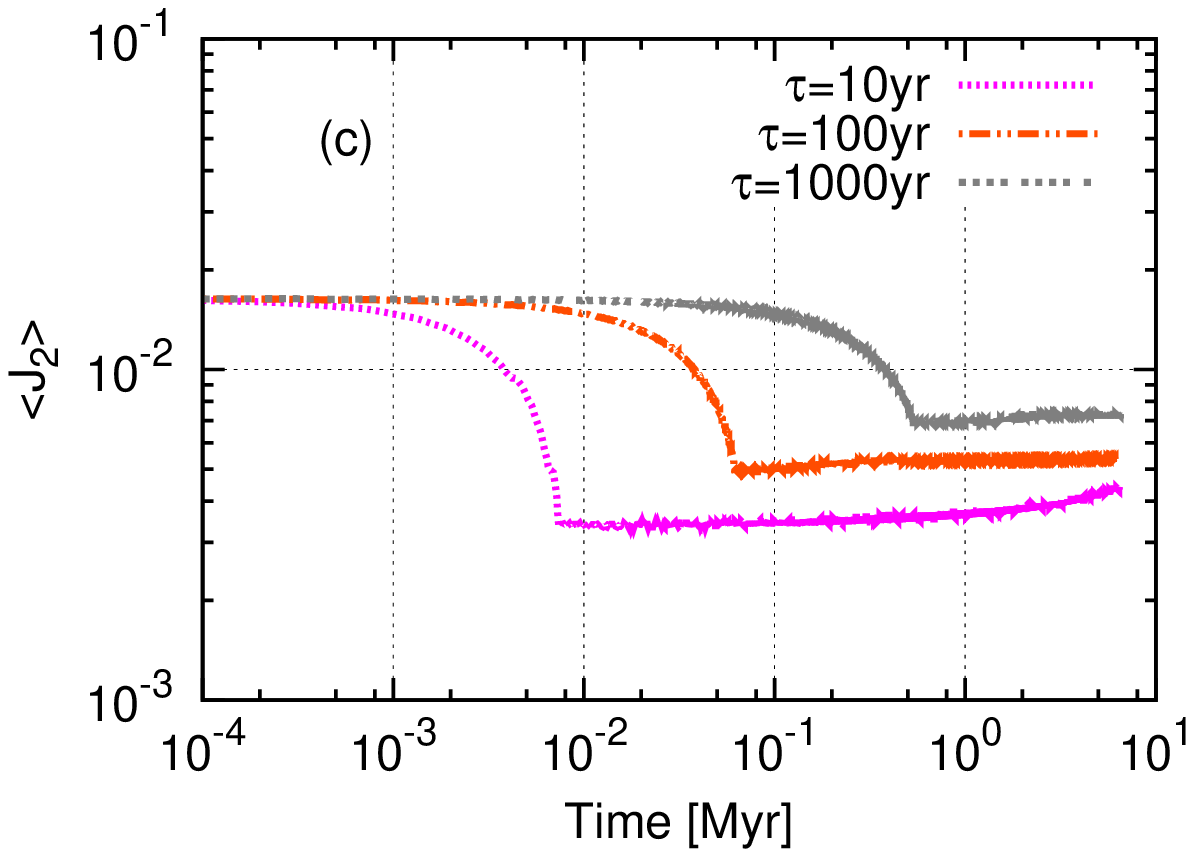}
\includegraphics[height=\columnwidth,angle=270]{\figpath 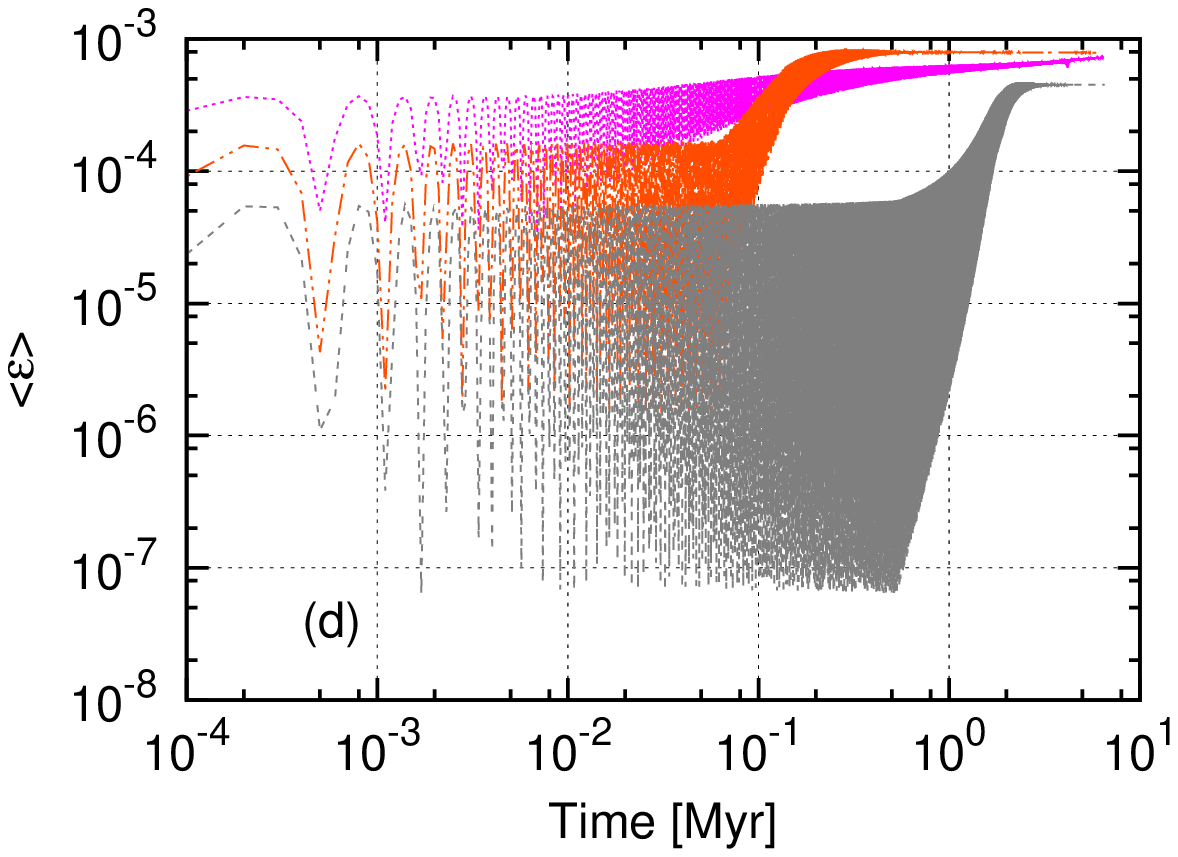}
\caption{\small Evolution of the shape of the planet with time for $\tau=10$\,yr,  $10^2$\,yr and $10^3$\,yr, and low initial eccentricity ($e_1=0.05$).
We show the instantaneous values of $J_2$ (a) and $\epsilon$ (b), together with their average values $\langle J_2\rangle$ (c) and $\langle \epsilon \rangle$ (d) (Eqs.\,(\ref{j2}) and (\ref{c22})).
The colors correspond to the same values of $\tau$ as in Fig.~\ref{eps}. 
The instantaneous values of $J_2$ and $\epsilon$ are still in agreement with the theoretical prediction for the averaged values.}
\label{c22-low}
\end{center}
\end{figure*}

In Fig. \ref{c22-low} we show the evolution of the instantaneous shape (top), together with \bfx{the corresponding analytical averaged} equilibrium values (bottom) given by expressions (\ref{j2}) and (\ref{c22}). 
As in the case with high initial eccentricity (section~\ref{hie}), the agreement between the numerical and the averaged deformation is very good. 
Note that, in all cases, the agreement for the prolateness begins when the rotation becomes trapped in the a SOR, because the averaged value
depends on $p$ (Eq. (\ref{c22})).
We conclude that expression (\ref{c22}) provides a good approximation for the shape of the body even when the orbit is excited by an external companion.

\section{Eccentricity pumping}
\label{pumping}

The initial secular increase observed for the eccentricity of the inner orbit (Figs.\,\ref{ecce-rot} and \ref{ecce-rot-low}) is \bfx{somewhat} unexpected, although a similar behaviour has already been described for gaseous planets \citep{Correia_etal_2012, Correia_etal_2013, Greenberg_etal_2013}.
In previous works, the eccentricity pumping is \bfx{related to} a variation in the $J_2$ of the inner planet due to the rotation, that tends to follow the pseudo-synchronous equilibrium (Eq.\,(\ref{090520a})). 
However, in the present case the rotation is always locked in a SOR, so this effect can be neglected.
In this section we show that the initial eccentricity pumping also results from a variation in $J_2$, but here the excitation directly comes from the tidal deformation with the adopted Maxwell rheology (Eq.\,\ref{max1})).

\subsection{Secular evolution of the eccentricity}

The eccentricity evolution of the inner orbit can be obtained from the Laplace-Runge-Lenz vector,
\begin{equation}
\vec{e}_1 = \frac{\dot{\vec r}_1 \times \vec{L}_1}{G m_0 m_1} -
\frac{\vec{r}_1}{r_1} 
\ , \label{100119a} 
\end{equation}
which points along the major axis in the direction of periapsis with magnitude $e_1 = |\vec{e}_1|$.
Thus
\begin{equation}
\dot{\vec e}_1 = \frac{1}{G m_0 m_1} \left( \vec{f}
\times {\vec{L}_1} + \dot{\vec r}_1 \times \dot{\vec L}_1 \right)
\label{100119b}  \ ,
\end{equation}
where $\vec{f}$ is the acceleration arising from the potential created by the deformation of the inner planet (Eq.\,(\ref{fdef})), and $\dot{\vec L}_1 = m_1 \, \vec{r}_1 \times \vec{f}$.
The secular evolution of the eccentricity can then be obtained by averaging over one orbital period 
\begin{equation}
\dot e_1 = \left\langle \frac{\dot{\vec e}_1 \cdot \vec{e}_1}{e_1} \right\rangle_{M_1} 
\ , \label{160324y}
\end{equation}
where $M_1$ is the mean anomaly of the inner planet's orbit.

For a single planet undergoing tidal dissipation with a Maxwell rheology, the secular evolution of the eccentricity becomes \citep[][Eq.\,(53)]{Correia_etal_2014}:
\begin{eqnarray}
\dot e_1 &=& - \frac{{\cal A}}{2} \left(\frac{R}{a_1} \right)^2 \frac{(1-e_1^2)}{e_1} \!\!   \sum_{k=-\infty}^{+\infty} \left[ \left(X_k^{-3,0}\right)^2 \! \frac{\tau k^2 n_1^2}{1+  \tau^2 k^2 n_1^2 } \right. \nonumber \\
&& + \left. \left(X_k^{-3,2}\right)^2 \frac{3 \tau \omega_k^2 }{1+  \tau^2 \omega_k^2} \left( 1 + \frac{2 n_1 / \omega_k}{\sqrt{1-e^2}}  \right) \right]   \ , 
\label{160323a} 
\end{eqnarray}
where $X_k^{l,m}$ is given by expression (\ref{061120gb}), and $ \omega_k = 2 \Omega - k n_1$.
The first term in expression (\ref{160323a}) results results from the contribution of the $J_2$, while the last term results from the contribution of $C_{22}$ and $S_{22}$ (see Eq.\,(\ref{fdef})).
When the rotation is captured in a SOR, $\Omega/n_1 = p = k/2$, it means that the tidal torque is dominated by the term with amplitude $X_{2p}^{-3,2}$, but also that $\omega_{2p} = 0$.
As a consequence, the eccentricity evolution is dominated by the $J_2$ contribution:
\begin{equation}
\dot e_1 \approx - \frac{{\cal A}}{2} \left(\frac{R}{a_1} \right)^2 \frac{(1-e_1^2)}{e_1} \!\!  \sum_{k=-\infty}^{+\infty} \left(X_k^{-3,0}\right)^2 \! \frac{\tau k^2 n_1^2}{1 +  \tau^2 k^2 n_1^2} 
\ . \label{160323b} 
\end{equation}
The coefficient $(X_k^{-3,0})^2$ is always positive and dominated by $e_1^{2k}$ \citep[e.g.][]{Laskar_Boue_2010}.
Since the term with $k=0$ is zero, the leading terms in the above series are for $k=\pm1$, and thus $\dot e_1 \propto - e_1$.
We hence conclude that the eccentricity of a single planet captured in a SOR can only decrease for a Maxwell rheology \citep[see][]{Correia_etal_2014}.

\subsection{Secular evolution of $J_2$}

In order to obtain expression (\ref{160323b}) we assumed a constant rotation rate and a constant eccentricity for $J_2^t$ (Eq.\,(\ref{max2})):
\begin{eqnarray}
J_2 &=& J_2^c + \frac{1}{\tau} \int_0^t J_2^t (t') \, \mathrm{e}^{(t'-t)/\tau} \, d t'
\nonumber \\ 
&=& J_2^c + {\cal A} \sum_{k=-\infty}^{+\infty} \frac{X_{k}^{-3,0}}{1 + \mathrm{i} \tau k n_1} \, \mathrm{e}^{\mathrm{i} k M_1} 
 \ , \label{160328c}
\end{eqnarray}
with $J_2^c = J_2^0 + J_2^r  = cte$.
When the rotation is captured in a SOR we can keep the assumption of constant rotation.
However, when the eccentricity is perturbed by a companion planet, the Hansen coefficients are no longer constant (Eq.\,(\ref{061120gb})).
Let us assume that the eccentricity is a periodic function with frequency $g$.
Thus,
\begin{equation}
X_k^{l,m} (e_1) =  \sum^{+\infty}_{j=-\infty} Y_{j,k}^{l,m} \, \mathrm{e}^{\mathrm{i} j g t} 
\ . \label{160323c}
\end{equation}
Using this expansion for $J_2^t$ (Eq.\,(\ref{max2})) in the computation of $J_2$ (Eq.\,(\ref{160328c})) gives:
\begin{equation}
J_2 = J_2^c + {\cal A} \sum_{j,k} \frac{Y_{j,k}^{-3,0}}{1 + \mathrm{i} \tau (k n_1 + j g)} \, \mathrm{e}^{\mathrm{i} (k n_1 + j g) t}
\ . \label{160323e} 
\end{equation}
In general, we have $g \ll n_1$.
Therefore, for all terms except $k=0$, we can neglect the contribution from $g$ in previous expression:
\begin{eqnarray}
J_2 \approx J_2^c 
&+& {\cal A} \sum_{k\ne0} \frac{X_k^{-3,0}}{1 + \mathrm{i} \tau k n_1} \, \mathrm{e}^{\mathrm{i} k n_1 t} 
\nonumber \\&+& {\cal A} \sum_{j} \frac{Y_{j,0}^{-3,0}}{1 + \mathrm{i} \tau j g} \, \mathrm{e}^{\mathrm{i} j g t} 
\ , \label{160323e} 
\end{eqnarray}
which gives
\begin{equation}
\langle J_2 \rangle \approx J_2^c + {\cal A} \sum_{j} \frac{Y_{j,0}^{-3,0}}{1 + \mathrm{i} \tau j g} \, \mathrm{e}^{\mathrm{i} j g t} \ .\label{160411d} 
\end{equation}
Noting also that 
\begin{equation}
{\cal A} \sum_{j} \frac{Y_{j,0}^{-3,0}}{1 + \mathrm{i} \tau j g} \, \mathrm{e}^{\mathrm{i} j g t} = \frac{1}{\tau} \int_0^t \langle J_2^t \rangle \, \mathrm{e}^{(t'-t)/\tau} \, d t'
\ , \label{160411c} 
\end{equation}
where $\langle J_2^t \rangle = {\cal A} X_0^{-3,0} (e)$ is the average value of $J_2^t$ over one orbital period (Eq.\,(\ref{j2e})),
we can thus obtain a secular version for the $\langle J_2 \rangle$ rheological law (Eq.\,(\ref{max1})) as 
\begin{equation}
\langle J_2 \rangle + \tau \langle \dot J_2 \rangle = J_2^c + \langle J_2^t \rangle 
\ . \label{160411b} 
\end{equation}

\subsection{Planetary perturbations}

We now consider the effect of the outer planet.
In absence of tidal deformation and dissipation, the eccentricity of the inner orbit is only perturbed by the outer companion.
Considering the leading orbital perturbations (octupole-level) we have \citep[e.g.][]{Correia_etal_2012}:
\begin{equation}
\dot e_1 \approx - \nu_{31}  \frac{e_2 (1 + 3/4 e_1^2) \sqrt{1-e_1^2}}{(1-e_2^2)^{5/2}} \sin \varpi  
\ , \label{110816h}
\end{equation}
and
\begin{eqnarray}
\dot \varpi \approx 
\frac{\nu_{gr}}{ (1-e_1^2)} 
+ \nu_{21} \frac{\sqrt{1-e_1^2}}{ (1-e_2^2)^{3/2}} - \nu_{22} \frac{ (1+\frac{3}{2} e_1^2)}{ (1-e_2^2)^2}  
 \ , \label{110819a}
\end{eqnarray}
with
\begin{equation}
\nu_{31} \approx n_1 \frac{15}{16} \frac{m_2}{m_0}  \left( \frac{a_1}{a_2} \right)^4 
\ , \quad
\nu_{gr} \approx 3 n_1  \left( \frac{n_1 a_1}{c} \right)^2
\ , \label{110817f}
\end{equation}
\begin{equation}
\nu_{21} \approx n_1 \frac{3}{4} \frac{m_2}{m_0}  \left( \frac{a_1}{a_2} \right)^3 
\ , \quad 
\nu_{22} \approx n_2 \frac{3}{4} \frac{m_1}{m_0} \left( \frac{a_1}{a_2} \right)^2
\ . \label{110817b}
\end{equation}
The variations in $e_2$ can be obtained from expression (\ref{am}).
The angle $\varpi = \varpi_1 - \varpi_2$ is the difference between the longitude of the periastron of the inner and outer orbits, and it can be also obtained from the Laplace vector (Eq.\,(\ref{100119b})) as
\begin{equation}
\dot \varpi_1 = \left\langle \dot{\vec e}_1 \cdot \left( \vec{k} \times \frac{\vec{e}_1}{e_1^2} \right) \right\rangle_{M_1}
\ . \label{160324x}
\end{equation}
In expression (\ref{110819a}) we only include the orbital perturbations (newtonian and general relativity corrections), but we also need to take into account the contribution from tides, given by $\vec{f}$ (Eq.\,(\ref{fdef})).
Considering only the leading $J_2$ term, we get
\begin{equation}
\dot \varpi_1 = \nu_1 \left[ \frac{\langle J_2 \rangle}{(1-e_1^2)^2} + {\cal A} F(e_1) \right]
\ , \label{160328a}
\end{equation}
with
\begin{equation}
\nu_1 = \frac{3 n_1}{2} \left(\frac{R}{a_1}\right)^2
\ , \label{160328b}
\end{equation}
and
\begin{equation}
F(e) = \frac{\sqrt{1-e^2}}{e} \sum_{k \ne 0} \frac{X_{k}^{-3,0}}{1 +  \tau^2 k^2 n_1^2} \left( X_{-k}^{-4,1}  + X_{-k}^{-4,-1}  \right)
\ . \label{160412b}
\end{equation}

\subsection{Linear approximation}

The complete secular evolution of the eccentricity of the inner orbit is given by the set of equations (\ref{160323b}), (\ref{160411b}), (\ref{110816h}), (\ref{110819a}) and (\ref{160328a}). 
Following \citet{Correia_etal_2012}, we can understand the unexpected increase of the eccentricity during the initial stages of the evolution by linearising the secular equations in the vicinity of the average values of $e_1$ and $J_2$.

Let $\langle J_2 \rangle = J_2^c + \delta J$, $e_1 = e_{10} + \delta e_1$ and $e_2 = e_{20} + \delta e_2$.
The $\delta e_2$ can be expressed as a function of $\delta e_1$ using the conservation of the orbital angular momentum (Eq.\,(\ref{am})), which we neglect since $\delta e_2 \ll \delta e_1$.
We also neglect the small damping effect given by expression (\ref{160323b}).
Then, assuming that $e_{i0} \ne 0$, the equations of motion (\ref{160411b}$-$\ref{110819a}, \ref{160328a}) reduce to:
\begin{equation}
\delta \dot e_1 = - \nu_e \sin \varpi \ , \label{110812b}
\end{equation}
\begin{equation}
\dot \varpi = g + g_e \delta e_1 + g_J \delta J \ , \label{110812c}
\end{equation}
\begin{equation}
\delta \dot J =  J_e \delta e_1 / \tau - \delta J / \tau
\ , \label{160411e} 
\end{equation}
with 
\begin{equation} 
\nu_e = \nu_{31} \frac{e_{20} (1 + 3/4 e_{10}^2) \sqrt{1-e_{10}^2}}{(1-e_{20}^2)^{5/2}} 
\ , 
\end{equation}
\begin{eqnarray} 
g &=& \frac{\nu_{gr}}{(1-e_{10}^2)} +\nu_1 \left[ \frac{J_2^c}{(1-e_{10}^2)^2} + {\cal A} f(e_{10}) \right]
\nonumber \\ &+& 
\nu_{21} \frac{\sqrt{1-e_{10}^2}}{(1-e_{20}^2)^{3/2}} - \nu_{22} \frac{(1+3 e_{10}^2/2)}{(1-e_{20}^2)^2} \ ,
\end{eqnarray}
\begin{eqnarray}
g_e &=& \frac{2 \nu_0 \, e_{10}}{(1-e_{10}^2)^2} + \nu_1 \left[ \frac{2 J_2^c \, e_{10}}{(1-e_{10}^2)^3} + {\cal A} \frac{\partial f}{\partial e} (e_{10}) \right]
\nonumber \\ &-&
\frac{\nu_{21} \, e_{10}}{\sqrt{1-e_{10}^2} (1-e_{20}^2)^{3/2}} 
 - \frac{3 \nu_{22} \, e_{10}}{(1-e_{20}^2)^2}  \ , 
\end{eqnarray}
\begin{equation}
g_J = \frac{\nu_1}{(1-e_{10}^2)^2}  \ , \label{160412a} 
\end{equation}
\begin{equation}
J_e =  \frac{3 {\cal A} \, e_{10}}{(1-e_{10}^2)^{5/2}}
\ . \label{160411f} 
\end{equation}

At first order, the precession of the periastron is constant
$ \dot \varpi \simeq g $, 
and the eccentricity is simply given from expression (\ref{110812b}) as
\begin{equation}
\delta e_1 = \Delta e \cos  (g t + \varpi_0) \ , \label{110812e}
\end{equation}
where $ \Delta e = \nu_e / g $, and $ \varpi =  g t + \varpi_0 $.
That is, the eccentricity $e_1$ presents periodic variations around an equilibrium value $ e_{10} $, with amplitude $ \Delta e $ and frequency $ g $.
Since $ g_e \delta e_1, g_J \delta J  \ll g$, the above solution for the eccentricity can be adopted as the zeroth order solution of the system of equations (\ref{110812b}$-$\ref{160411e}).
With this approximation, the equation of motion of $\delta J$ (\ref{160411e}) becomes that of a driven harmonic oscillator whose steady state solution is
\begin{equation}
\delta J = \Delta J \cos ( g t + \varpi_0 - \phi) \ , \label{110920a}
\end{equation}
with 
\begin{equation}
\Delta J = \frac{J_e \Delta e}{\sqrt{1 + (\tau g)^2} } 
\ , \quad \mathrm{and} \quad
\sin \phi = \frac{\tau g}{\sqrt{1 + (\tau g)^2}}
\ . \label{160412c} 
\end{equation}
The $J_2$ thus presents an oscillation identical to the eccentricity (Eq.(\ref{110812e})), but delayed by an angle $\phi$.
Using the above expression in equation (\ref{110812c}) and integrating, gives for the periastron:
\begin{equation}
\varpi = g t + \varpi_0 + \frac{g_e}{g} \Delta e \sin (g t + \varpi_0) + \frac{g_J}{g} \Delta J \sin (g t + \varpi_0 - \phi)
\label{110812g} \ .
\end{equation}
Finally, substituting in expression (\ref{110812b}) and using the approximation $g_e \Delta e, g_J \Delta J \ll g$ gives
\begin{eqnarray}
\delta \dot e_1 
& \approx & - \nu_e \sin (g t + \varpi_0) \nonumber \\ 
& & - \nu_e \frac{g_e}{g} \Delta e \sin (g t + \varpi_0) \cos  (g t + \varpi_0) \nonumber \\ 
& & - \nu_e \frac{g_J}{g} \Delta J \sin (g t + \varpi_0 - \phi) \cos  (g t + \varpi_0) 
  \label{110812h} \ ,
\end{eqnarray}
or, combining the two products of periodic functions,
\begin{eqnarray}
\delta \dot e_1  & = & - \nu_e \sin (g t + \varpi_0)  +  \nu_e \frac{g_J}{2 g} \Delta J \sin \phi \nonumber \\
& & - \nu_e \frac{g_e}{2 g} \Delta e \sin (2 g t + 2 \varpi_0)  \nonumber \\
& & - \nu_e \frac{g_J}{2 g} \Delta J \sin (2 g t + 2 \varpi_0 - \phi)\label{110812i} \ .
\end{eqnarray}
The last two terms in previous equation can be neglected since they are
periodic and have a very small amplitude ($g_e \Delta e, g_J \Delta J \ll g$). 
However, the second term in $\sin \phi$ is constant and it adds an increasing drift to the eccentricity,
\begin{equation}
\langle \delta \dot e_1 \rangle = \nu_e \frac{J_e \Delta e}{2} \frac{\tau g_J}{1 + (\tau g)^2} 
 \ . \label{110902b}
\end{equation}
The drift is maximized for $ \tau g \sim 1 $, which corresponds to $\phi \sim 45^\circ$ (Eq.\,(\ref{160412c})).  
It vanishes for weak dissipation ($\tau g \ll 1 $), but also for strong dissipation ($\tau g \gg 1 $).
The phase lag $\phi$ between the eccentricity (Eq.\,(\ref{110812e})) and the $J_2$ variations (Eq.\,(\ref{110920a})) is thus essential to get a drift on the eccentricity.
The eccentricity pumping was never observed in previous studies with visco-elastic rheologies, since tidal deformation and dissipation are given in the Fourier domain by the complex Love number $k_2$ (which is computed for a given frequency), while here we use a time-dependent rheological law (Eq.\,(\ref{max1})) that allows \bfx{this kind of} feedback effects.

The major difference when we consider the full non-linearized problem
is that the drift (Eq.\,(\ref{110902b})) cannot grow indefinitely.
Indeed, when the eccentricity reaches high values, the drift vanishes (Fig.\,\ref{ecce-rot-low}).
Moreover, the tidal damping of the eccentricity is also enhanced for high eccentricities (Eq.\,(\ref{160323b})), which counterbalances the drift \bfx{(Fig.\,\ref{ecce-rot})}.
Although the pumping drift can be present for the age of the system, when the amplitude of the eccentricity oscillations becomes small ($\Delta e \rightarrow 0$), the drift disappears (Eq.\,(\ref{110902b})) and the eccentricity can only be damped.

\section{Discussion and conclusions}
\label{disc}

In this paper we have studied the coupled orbital and spin evolution of the CoRoT-7 two-planet system using a Maxwell viscoelastic rheology for the inner planet.
This rheology is characterized by a viscous relaxation time, $\tau$, that can be seen as the characteristic average time that the planet requires to achieve a new equilibrium shape after being disturbed by an external forcing.

We studied the past evolution of the system adopting different values for the relaxation time of CoRoT-7\,b, ranging from a few hours up to one century ($10^{-3}-10^2$\,yr).
In all situations, the spin evolves quickly until it is captured in some SOR.
It then follows through a successive temporarily trappings in SORs, which are progressively destabilized as the eccentricity decays.
\bfx{Several works on tidal evolution usually assume synchronous motion for the rotation of the close-in companions, as this is the natural outcome resulting from tidal interactions.
Nevertheless, for large values of the relaxation times, which is likely the case for most terrestrial planets, we note that the rotation can remain trapped into high-order SORs for tens of Myr.} 

We observed that there are two different regimes for the orbital evolution.
For small $\tau$ values (0.01$-$0.1\,yr), the eccentricity of both orbits is rapidly damped, in agreement with previous results \citep[e.g.,][]{Ferraz-Mello_etal_2011, Rodriguez_etal_2011, Dong_Ji_2012}. 
However, for large $\tau$ values ($10-10^2$\,yr), the inner planet eccentricity is pumped to higher values, whereas the outer planet eccentricity is simultaneously damped due to the orbital
angular momentum conservation.

The inner orbit eccentricity pumping 
was already reported in previous works that used the linear model instead of the Maxwell one \citep{Correia_etal_2012, Correia_etal_2013, Greenberg_etal_2013}.
In these works, the effect resulted from a forced excitation of the $J_2$ due to oscillations in the rotation rate.
This mechanism works as long as the rotation is close to the pseudo-synchornous state and undergoes variations due to the eccentricity forcing \citep[see][]{Correia_2011}.
Although the pseudo-synchornous state can be expected for gaseous planets, for rocky planets the spin always ends up trapped in a SOR due to the permanent equatorial deformation.
Thus, for this class of planets, the pumping mechanism identified by \citet{Correia_etal_2012} does not work.

The eccentricity pumping described in this paper also results from a forced excitation of the $J_2$ of the planet, but due to the tidal deformation.
Indeed, the equilibrium $J_2$ has a rotational (Eq.\,(\ref{j2r})) and a tidal contribution (Eq.\,(\ref{max2})), but inside a SOR the rotational contribution is nearly constant, while the tidal one still undergoes variations due to the term in $r_1^{-3}$.
The pumping effect is an important mechanism that may help to explain the non-zero eccentricity presently observed for the orbit of CoRoT-7\,b.

Due to the computational cost of the numerical simulations, we were not able to perform here a large set of runs for different planetary systems. 
However, we have shown that at least for the CoRoT-7 system unexpected behaviors can occur when we take into account the coupled orbital and spin evolution.
\bfx{In particular, the non-zero eccentricities observed for many other close-in super-Earths in multiple planetary systems, may be explained by similar pumping mechanisms.}

Since the Maxwell model is more realistic than the constant$-Q$ and the constant time lag models, the results described in this paper provide a more accurate picture for the diversity of behaviors among planetary systems that interact by tides.
Alternative viscoelastic rheologies to the Maxwell model exist, such as the Standard Anelastic Solid model \citep[e.g.,][]{Henning_etal_2009} or the Andrade model \citep[e.g.,][]{Efroimsky_2012}.
These rheologies may also be able to reproduce the pumping effect on the inner orbit eccentricity.
Note, however, that in order to observe the excitation in $J_2$ due to the eccentricity forcing, we need to use a time-dependent rheological law similar to expression (\ref{max1}) that allows feedback effects.

\bfx{In this study we considered coplanar orbits and the spin of the planet orthogonal to the orbits (zero obliquity).
Although multi-planet systems usually present low mutual inclinations of about $1^\circ$ on average \citep{Figueira_etal_2012, Tremaine_Dong_2012}, this value can be large enough to perturb the long-term evolution of the obliquity \citep{Laskar_Robutel_1993, Correia_Laskar_2003I}.
Our model can be easily extended to non-planar configurations (for planets with some obliquity and evolving in inclined orbits), provided that we additionally take into account the deformation of the $C_{21}$ and $S_{21}$ gravity field coefficients in the gravitational potential, as explained in \citet{Boue_etal_2016}.}



\subsection*{Acknowledgments}
We acknowledge support from FAPESP  (2013/16771-6 and 2013/21891-0) and from CIDMA strategic project UID/MAT/04106/2013.

\bibliographystyle{mnras}      
\bibliography{references}   

\end{document}